\documentclass[preprint,12pt]{elsarticle}

\usepackage{float}
\usepackage{graphicx}
\usepackage{subcaption}
\usepackage{amssymb}
\usepackage{amsmath}
\usepackage{booktabs}
\usepackage{multirow} 
\usepackage{tabularx}
\usepackage{rotating, rotfloat} % enables [!t] on sidewaysfigure*

\usepackage{pdflscape} % for landscape environment
\usepackage{tikz}
\usetikzlibrary{positioning,arrows.meta,shapes.geometric}
\usepackage[utf8]{inputenc}
\journal{Actia Materialia}

\begin{document}

\begin{frontmatter}

\title{Sequential Bayesian Inference of the GTN Damage Model Using Multimodal Experimental Data}

\author[inst1]{Mohammad Ali Seyed Mahmoud}
\author[inst1]{Dominic Renner}
\author[inst3]{Ali Khosravani}
\author[inst1,inst2]{Surya R. Kalidindi}

% Affiliation 1 (already in your file)
\affiliation[inst1]{%
  organization={George W. Woodruff School of Mechanical Engineering, Georgia Institute of Technology},
  addressline={North Avenue NW},
  city={Atlanta},
  state={GA},
  country={USA}
}

% Affiliation 2 (your text)
\affiliation[inst2]{%
  organization={School of Computational Science and Engineering, Georgia Institute of Technology},
  city={Atlanta},
  state={GA},
  country={USA}
}

% Affiliation 3 (your text)
\affiliation[inst3]{%
  organization={Exponent},
  city={Menlo Park},
  state={CA},
  country={USA}
}
\begin{abstract}
Reliable parameter identification in ductile damage models remains challenging because the salient physics of damage progression are localized to small regions in material responses, and their signatures are often diluted in specimen-level measurements. Here, we propose a sequential Bayesian Inference (BI) framework for the calibration of the Gurson-Tvergaard-Needleman (GTN) model using multimodal experimental data (i.e., the specimen-level force-displacement (F-D) measurements and the spatially resolved digital image correlation (DIC) strain fields). This calibration approach builds on a previously developed two-step BI framework that first establishes a low-computational-cost emulator for a physics-based simulator (here, a finite element model incorporating the GTN material model) and then uses the experimental data to sample posteriors for the material model parameters using the Transitional Markov Chain Monte Carlo (T-MCMC). A central challenge to the successful application of this BI framework to the present problem arises from the high-dimensional representations needed to capture the salient features embedded in the F-D curves and the DIC fields. In this paper, it is demonstrated that Principal Component Analysis (PCA) provides low-dimensional representations that make it possible to apply the BI framework to the problem. Most importantly, it is shown that the sequence in which the BI is applied has a dramatic influence on the results obtained. Specifically, it is observed that applying BI first on F-D curves and subsequently on the DIC fields produces improved estimates of the GTN parameters. Possible causes for these observations are discussed in this paper, using AA6111 aluminum alloy as a case study.
\end{abstract}

\begin{graphicalabstract}
\includegraphics[width=\textwidth]{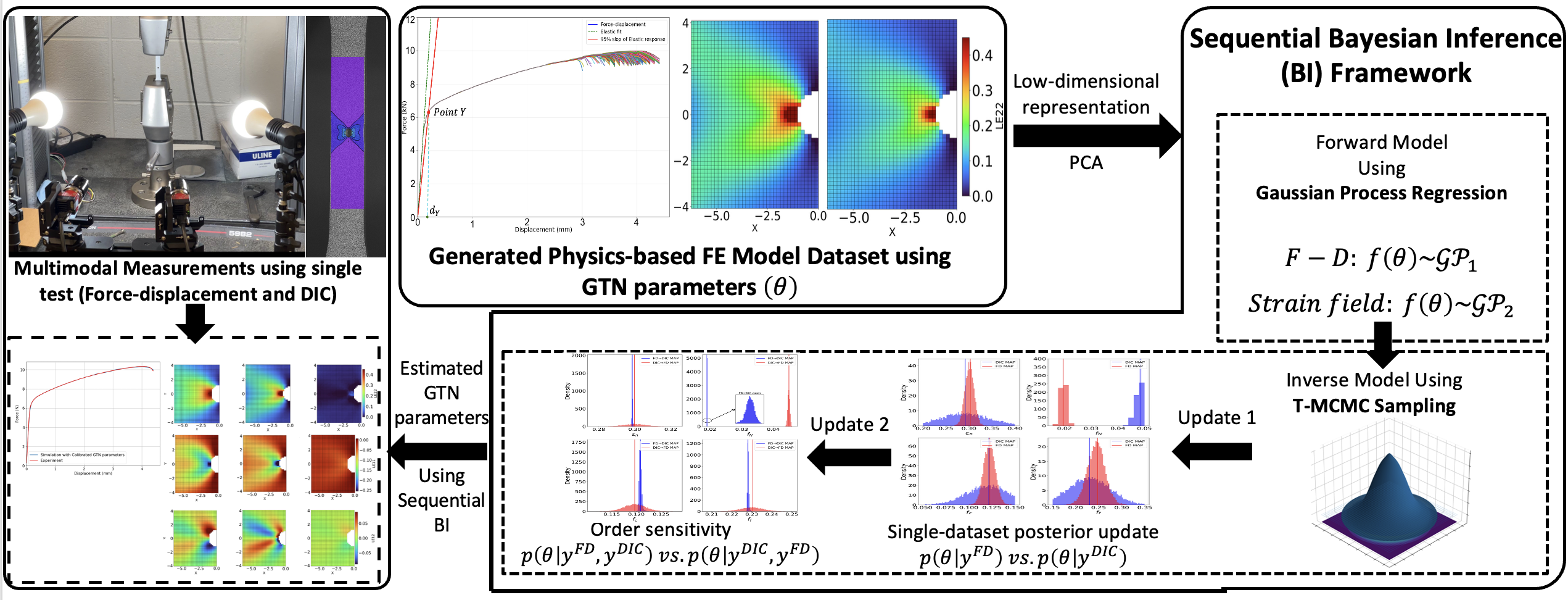} 
\end{graphicalabstract}


\end{frontmatter}

\section{Introduction}
\label{sec:introduction}

Accurate estimation of constitutive model parameters is essential for reliable predictions of material behavior in almost all engineering applications~\citep{Generale2024InverseMS,Kalidindi2022DT, Marshall2021JOM,Daviran2024Materialia}. Conventional approaches for material parameter estimation typically employ deterministic, gradient-descent based optimization techniques to minimize the discrepancy between measured experimental responses (e.g., stress–strain curves) and the corresponding model predictions (analytically derived or numerically computed using finite element (FE) models) \citep{tarantola2005inverse, walter1997identification}. Such deterministic approaches generally provide point estimates that are most likely to represent local minima rather than global optima \citep{beck1998updating, kaipio2005statistical, Mahmoud2023Materials}. Furthermore, the conventional optimization procedures do not explicitly incorporate or quantify the inherent uncertainties arising from experimental measurements \citep{Venkatraman2022}. These uncertainties, coupled with fundamental identifiability issues (i.e., multiple parameter combinations can yield equally plausible fits), compromise the reliability of the estimated parameter values \citep{Generale2022}. Recent advances in the use of Bayesian Inference (BI) for parameter estimation have shown significant promise in addressing the challenges described above. Specifically, a two-step BI framework~\citep{Castillo2019Frontiers,Castillo2019JOM,FernandezZelaia2018} has demonstrated tremendous promise in addressing the core identifiability issue encountered in calibrating material models with large numbers of fittable parameters, especially when there is a limited amount of available experimental data \citep{Generale2022,Venkatraman2025, castillo2021bayesian, Ray2025npjCM}. Moreover, such probabilistic approaches enable systematic incorporation of prior knowledge and experimental uncertainty \citep{Venkatraman2025}.

Parameter estimation becomes significantly more complex when applied to sophisticated constitutive descriptions such as the Gurson-Tvergaard-Needleman (GTN) model~\citep{Tvergaard1984}, which aims to describe damage evolution and failure in ductile metals. Experiments studying damage evolution typically involve heterogeneous deformations with steep strain gradients (e.g., necking in tensile tests), necessitating advanced data analysis protocols \citep{Vogler2008, Cao2014, Bron2002}. Furthermore, several of the parameters involved in damage models cannot be measured directly from experiments. For example, the GTN model requires characterization of microstructural parameters describing the initial void volume fraction (representing pre-existing porosity in the undeformed material), critical void fraction at the onset of coalescence (marking the transition from isolated void growth to coalescence-driven damage), and void interaction parameters (governing the evolution of void spacing and the resulting influence on stress states) \citep{Leblond2014, Zhang2022}. Direct measurement of such microstructural parameters is often unreliable due to the extensive imaging and sophisticated image analyses required to accurately quantify these variables in any given sample \citep{Maire2014, Depraetere2022}. Consequently, indirect calibration methods that align model predictions with experimentally measured material responses become essential. 
One further complication in the estimation of material parameters encountered in damage models such as the GTN model stems from the heavy reliance on averaged bulk response (e.g., F-D curves). This approach inherently smears out any observations of the critical localized deformation effects fundamental to damage progression \citep{Roux2015, Avril2008}.

Recent advances in Digital Image Correlation (DIC) \citep{Sutton2009,Pan2009,Mousa2023,Li2024} have enabled direct measurement of heterogeneous strain fields at damage initiation sites, unleashing new avenues for calibrating complex constitutive models such as the GTN ductile damage model \citep{Mousa2023, Gerbig2016}. Specifically, a recent study has explored the calibration of constitutive models with DIC measurements through Finite Element Model Updating (FEMU), which involves adjusting model parameters to minimize discrepancies between experimental and simulated strain fields \citep{Chen2025, Gerbig2016}. These FEMU-based approaches typically employ grid-alignment procedures to match experimental DIC measurement points with finite-element computational nodes, and employ error metrics based on the L2-norm of the pointwise strain differences \citep{Fayad2022}. However, several challenges remain unresolved. First, the high dimensional full-field DIC data complicates the identification of the most informative spatial features for robust parameter estimation \citep{Mokhtarian2013, Pont2025}. Second, mismatches between experimental measurement grids and computational mesh discretization introduce additional difficulties, creating computationally intensive and ill-conditioned inverse problems that compromise the reliability of parameter estimates \citep{Fayad2022, Mokhtarian2013}. Third, current FEMU implementations have been demonstrated primarily for relatively simpler constitutive models, such as anisotropic yield surfaces (Hill’s 1948 yield criterion) and Johnson–Cook plasticity models \citep{Lattanzi2020, Pathirane2021}. They have not yet been successfully applied to damage models such as the GTN model. Finally, most previous studies have relied on a single form of experimental data, employing F-D curves or DIC strain fields, but not both.

In this study, we present a methodological innovation through the extension of the two-step BI framework to utilize multimodal data streams. Specifically, we develop the necessary protocols for successfully obtaining useful posteriors on GTN model parameters while using both bulk F-D measurements and the DIC-measured localized strain fields. Due to the inherent complexity associated with constructing a single joint likelihood that encompasses both global F-D and full-field strain observations, a sequential Bayesian updating approach becomes necessary and useful. The ensuing sequential nature of posterior updating in this framework requires an investigation into how the order of data assimilation influences parameter identification results. Although the theory suggests that posterior distributions extracted from BI should be independent of the update sequence, practical implementations often exhibit significant sensitivity to the sequencing order. For instance, in structural health monitoring \citep{Li2024}, updating particle-filter models with high-precision accelerometer data before displacement measurements has been shown to cause sample collapse and biased parameter estimates, whereas reversing the sequence preserves accuracy. Similarly, geophysical reservoir characterization studies demonstrate that sequential importance sampling produces markedly different parameter posteriors when sparse pressure logs precede information-rich seismic data, attributed to unequal conditioning between likelihood functions \citep{Iglesias2013}. These studies collectively highlight the importance of update order. This research aims to demonstrate the benefits of the sequential BI framework for the reliable estimation of the GTN model parameters using multimodal experimental data. It also critically evaluates the relative informational contributions of the distinct data streams and the influence of the update sequence on the estimated values of the parameters.

\section{Background}
\label{sec:theoretical_background}

This section summarizes the mathematical background underlying the different methods employed within this work, beginning with the GTN damage model and ending with the BI methods used for parameter estimation.

\subsection{GTN Ductile Damage Model}
\label{sec:gtn_model}

The Gurson-Tvergaard-Needleman (GTN) model provides a micro-mechanical framework to describe ductile damage evolution through void nucleation, growth, and coalescence mechanisms~\citep{Tvergaard1984,Gurson1977,Needleman1987,Benzerga2004}. It utilizes an augmented form of the von Mises yield criterion by accounting for the presence and evolution of porosity and its effects, and is expressed as:

\begin{equation}
\Phi(\sigma, f^*) = \left(\frac{\sigma_{\text{eq}}}{\sigma_y}\right)^2 + 2q_1 f^* \cosh\left(\frac{3q_2\sigma_m}{2\sigma_y}\right) - \left[1 + q_3 {f^*}^2\right] = 0,
\label{eq:gtn_yield}
\end{equation}

\noindent where $\sigma_{\text{eq}}$, $\sigma_m$, and $\sigma_y$ represent the equivalent von Mises stress, hydrostatic stress, and the matrix (solid continua) yield strength, respectively. The effective void fraction \( f^* \) captures augmented porosity effects during coalescence, and is defined as

\begin{equation}
f^* = \begin{cases}
f & f < f_c,\\
f_c + \left(\frac{1}{q_1} - f_c\right)\frac{f - f_c}{f_f - f_c}, & f \ge f_c,
\end{cases}
\label{eq:effective_f}
\end{equation}

\noindent where $f_c$ and $f_f$ denote the coalescence and fracture thresholds, respectively~\citep{Needleman1987}. The coefficients $q_1$, $q_2$, and $q_3$ in Eq.~(\ref{eq:gtn_yield}) modulate the void interaction and shape effects~\citep{Tvergaard1984,Hardin2011}.

Void evolution combines growth and nucleation, expressed as

\begin{align}
\dot{f} &= \dot{f}_g + \dot{f}_n
\label{eq:void_evolution}\\
\dot{f}_g &= (1 - f) \mathrm{tr}\left(\dot{\boldsymbol{\varepsilon}}^p\right) \label{eq:void_growth}\\
\dot{f}_n &= \frac{f_N}{S_N \sqrt{2\pi}} \exp\left[-\frac{(\varepsilon^p - \varepsilon_N)^2}{2S_N^2}\right] \dot{\varepsilon}^p 
\label{eq:void_nucleation}
\end{align}

\noindent where growth is driven by plastic dilatation~\citep{Rice1969} and nucleation is modeled via a normal distribution centered at $\varepsilon_N$ with standard deviation $S_N$~\citep{Chu1980}. Here, $f_N$ quantifies the total volume fraction available for nucleation.

\subsection{Bayesian Inference Framework For Parameter Estimation}
\label{subsec:bayesian_framework}

 Bayesian Inference (BI) provides a probabilistic approach to parameter estimation that systematically incorporates prior knowledge and experimental uncertainty while quantifying parameter correlations and confidence bounds \citep{Beck1998, Venkatraman2025}. This framework treats model parameters as random variables with associated probability distributions, enabling rigorous uncertainty quantification in parameter estimation \citep{Rappel2019}. The fundamental principle of BI for updating prior beliefs about model parameters $\boldsymbol{\theta}$ given experimental evidence $\mathbf{y}$ is expressed as
\begin{equation}
p(\boldsymbol{\theta}|\mathbf{y}) = \frac{p(\mathbf{y}|\boldsymbol{\theta}) p(\boldsymbol{\theta})}{p(\mathbf{y})}
\label{eq:bi}
\end{equation}
where $p(\boldsymbol{\theta}|\mathbf{y})$ represents the model parameter posterior, $p(\mathbf{y}|\boldsymbol{\theta})$ is the likelihood, $p(\boldsymbol{\theta})$ represents the prior knowledge of the model parameters, and $p(\mathbf{y})$ is the evidence. The posterior encapsulates updated information about possible combinations of parameter values along with their uncertainties and interdependencies (i.e., correlations) \citep{Gelman2013}. In practice, Eq.~(\ref{eq:bi}) can be implemented using a two-step approach, described next.

\subsubsection{Two-Step Bayesian Inference for Material Model Parameter Estimation}
The first hurdle encountered in the implementation of Eq.~(\ref{eq:bi}) comes from the fact that the direct evaluation of the likelihood term, $p(\mathbf{y}|\boldsymbol{\theta})$, often requires a very large number of expensive computations. In the context of this paper, this requires the computation of the overall sample response for each specified set of material parameter values, often using FE simulations. Gaussian Process (GP) regression provides an efficient surrogate modeling approach that approximates the relationship between the GTN model parameters and their corresponding responses while quantifying prediction uncertainty~\citep{Rasmussen2006}. The establishment of this surrogate model (also referred to as an emulator of the physics-based simulations) constitutes the first step of the two-step BI framework employed in this work. Note that this step can be completed without any experimental data. One would perform an adequate number of simulations covering the full range of the parameter space of interest. 

A GP is completely specified by its mean function $m(\boldsymbol{\theta})$ and covariance function $k(\boldsymbol{\theta}, \boldsymbol{\theta}')$ as
\begin{equation}
f(\boldsymbol{\theta}) \sim \mathcal{GP}(m(\boldsymbol{\theta}), k(\boldsymbol{\theta}, \boldsymbol{\theta}'))
\end{equation}
The GP framework treats the unknown function as a realization of a stochastic process, and enables probabilistic predictions at arbitrary input values \citep{Kennedy2001}. GPs are usually regressed to training data (in our case produced using FE simulations). We quantify the probability of observing the experimental data, given model parameters and the trained GP surrogate, via the following likelihood. For multimodal experimental data, separate GPs are trained for each data stream because the outputs differ, even though the inputs are the same.

The likelihood for any data stream takes the form:
\begin{equation}
p(\mathbf{y}\mid\boldsymbol{\theta})
= \prod_{i=1}^{N} \frac{1}{\sqrt{2\pi\sigma^2_i}}
\exp\!\left(-\frac{\big(y_i - f_i(\boldsymbol{\theta})\big)^2}{2\sigma^2_i}\right)
\end{equation}
\noindent where $N$ denotes the number of data points, $y_i$ represents the $i$-th experimental observation, $f_i(\boldsymbol{\theta})$ is the corresponding model prediction from the GP surrogate, and $\sigma_i$ represents the measurement uncertainty for the $i$-th data point. This formulation assumes independent Gaussian measurement errors, a standard assumption that facilitates tractable posterior computation. Noise levels are determined from instrument specifications and validation procedures: $\sigma_{\mathrm{FD}}$ from load-cell calibration protocols and $\sigma_{\mathrm{DIC}}$ from DIC system characterization studies (see Section~\ref{sec:experiments}).

The second step of the two-step BI framework focuses on sampling from the posterior expressed in Eq.~(\ref{eq:bi}). We resort to sampling the posterior because analytical evaluation of the posterior distribution is often intractable. Markov Chain Monte Carlo (MCMC) methods provide a robust framework for generating samples from posteriors~expressed in Eq.~(\ref{eq:bi}) \citep{Brooks2011}. In this work, Transitional Markov Chain Monte Carlo (T-MCMC) is employed to gradually transform the prior to the posterior through intermediate distributions, improving sampling efficiency in complex likelihood landscapes \citep{Ching2007}. Convergence assessment through diagnostics such as the Gelman-Rubin statistic and effective sample size calculations ensures reliable posterior characterization and parameter uncertainty quantification \citep{Gelman1992}.

\subsubsection{Sequential multimodal updating}

\begin{figure}[htbp]
\centering
\begin{tikzpicture}[
    node distance=2cm,
    box/.style={rectangle, draw, thick, rounded corners, minimum width=4cm, minimum height=1cm, align=center},
    arrow/.style={->, thick, blue}
]

% Sequence 1
\node [box, fill=blue!20] (seq1_title) at (0,4) {\textbf{Sequence 1:}\\F-D → DIC};
\node [box, fill=orange!10] (seq1_step1) at (0,2.5) {\textbf{Update 1:}\\ Two-step BI Using F-D data};
\node [box, fill=red!10] (seq1_step2) at (0,0.5) {\textbf{Update 2:}\\Two-step BI Using DIC data \\ (F-D Posterior as Prior)};

% Sequence 2  
\node [box, fill=red!20] (seq2_title) at (8,4) {\textbf{Sequence 2:} \\DIC → F-D};
\node [box, fill=orange!10] (seq2_step1) at (8,2.5) {\textbf{Update 1:}\\Two-step BI Using DIC data};
\node [box, fill=red!10] (seq2_step2) at (8,0.5) {\textbf{Update 2:}\\Two-step BI Using F-D data \\ (DIC Posterior as Prior)};

% Arrows
\draw [arrow] (seq1_step1) -- (seq1_step2);
\draw [arrow] (seq2_step1) -- (seq2_step2);

% Comparison
\node [box, fill=green!10] (comparison) at (4,-2) {Compare Final Posteriors};
\draw [arrow, dashed] (seq1_step2) to [out=270, in=135] (comparison);
\draw [arrow, dashed] (seq2_step2) to [out=270, in=45] (comparison);

\end{tikzpicture}
\caption{Sequential Bayesian Inference framework comparing two update sequences. Each sequence provides both final posteriors and intermediate single-update results for evaluating individual data stream contributions to GTN parameter estimation.}
\label{fig:methodology_simple}
\end{figure}
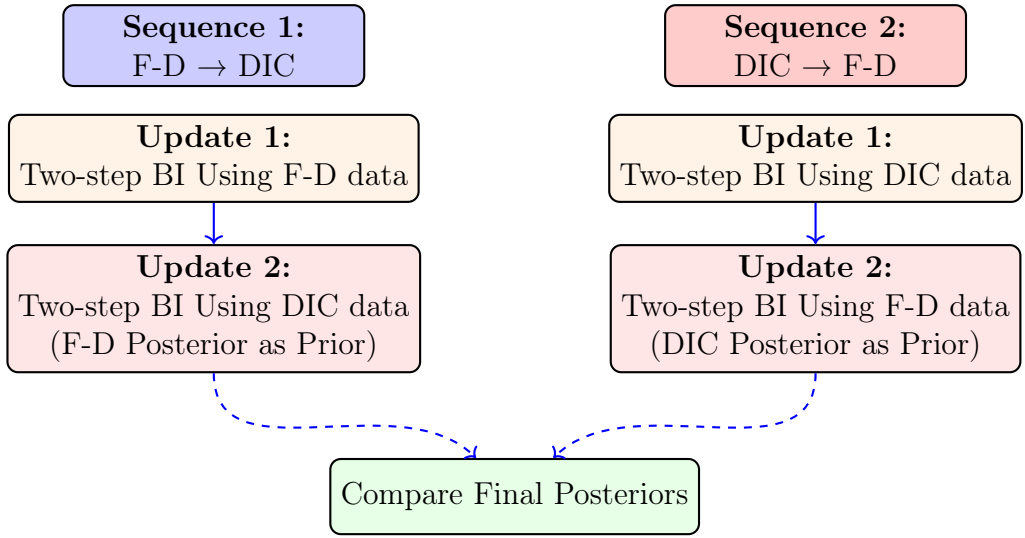

As already discussed, practical implementation of sequential BI can be order-sensitive~\citep{Iglesias2013}. To critically examine this phenomenon in the present application, we explore the following two sequences (see Figure~\ref{fig:methodology_simple}):

\noindent\textbf{Sequence 1}: Update with F-D measurement, then with DIC measurement (F-D$\rightarrow$ DIC).
\begin{equation}
  p(\boldsymbol{\theta}\mid \mathbf{y}^{\mathrm{FD}}) \propto p(\mathbf{y}^{\mathrm{FD}}\mid \boldsymbol{\theta})\,p(\boldsymbol{\theta})
\end{equation}

$$\qquad
  p(\boldsymbol{\theta}\mid \mathbf{y}^{\mathrm{FD}}, \mathbf{y}^{\mathrm{DIC}}) \propto p(\mathbf{y}^{\mathrm{DIC}}\mid \boldsymbol{\theta})\,p(\boldsymbol{\theta}\mid \mathbf{y}^{\mathrm{FD}})$$

\noindent\textbf{Sequence 2}: Update with DIC measurement, then with F-D measurement (DIC$\rightarrow$ F-D).

\begin{equation}
  p(\boldsymbol{\theta}\mid \mathbf{y}^{\mathrm{DIC}}) \propto p(\mathbf{y}^{\mathrm{DIC}}\mid \boldsymbol{\theta})\,p(\boldsymbol{\theta}) 
\end{equation}

$$\qquad
  p(\boldsymbol{\theta}\mid \mathbf{y}^{\mathrm{DIC}},\mathbf{y}^{\mathrm{FD}}) \propto p(\mathbf{y}^{\mathrm{FD}}\mid \boldsymbol{\theta})\,p(\boldsymbol{\theta}\mid \mathbf{y}^{\mathrm{DIC}})
$$

\noindent Note that these two  sequences also provide information on single updates \(p(\boldsymbol{\theta}\mid \mathbf{y}^{\mathrm{FD}})\) and \(p(\boldsymbol{\theta}\mid \mathbf{y}^{\mathrm{DIC}})\), and enable critical evaluation of the value of each data stream for GTN parameter estimation.

\section{Experiments}
\label{sec:experiments}

This section outlines the experimental protocols employed to simultaneously capture both the specimen-level F-D responses and full-field DIC measurements. These measurements provide the multimodal dataset required for GTN model parameter estimation using the sequential BI framework.

\subsection{Material Selection and Specimen Design}
\label{subsec:material_design}

The experiments utilized AA6111-T4 aluminum alloy, selected for its well-documented ductile behavior and suitability for damage model validation \citep{Zheng2003, Li2005, Dunand2012, Moradi2022IJLMM}. This precipitation-hardened Al-Mg-Si alloy exhibits an optimal balance of strength and formability in the T4 temper condition, ensuring reliable representation of deformation behavior including void nucleation and damage evolution. The alloy's chemical composition, conforming to standard specification ranges (Table~\ref{tab:composition}), provides consistent material properties essential for reproducible calibration results.

\begin{table}[htbp]
\centering
\caption{Chemical composition of AA6111-T4 aluminum alloy (wt\%).}
\label{tab:composition}
\begin{tabular}{lcc}
\hline
Element & Composition (wt\%) & Specification Range \\
\hline
Aluminum (Al)  & Balance      & --- \\
Magnesium (Mg) & 0.4--0.8    & 0.4--0.8 \\
Silicon (Si)   & 0.2--0.6    & 0.2--0.6 \\
Copper (Cu)    & 0.05--0.1   & 0.05--0.1 \\
Manganese (Mn) & 0.05--0.2   & 0.05--0.2 \\
Iron (Fe)      & 0.1--0.3    & 0.1--0.3 \\
Titanium (Ti)  & 0.05--0.15  & 0.05--0.15 \\
\hline
\end{tabular}
\end{table}

\begin{figure}[htbp]
    \centering
    % --- first image ---
    \includegraphics[width=0.5\textwidth]{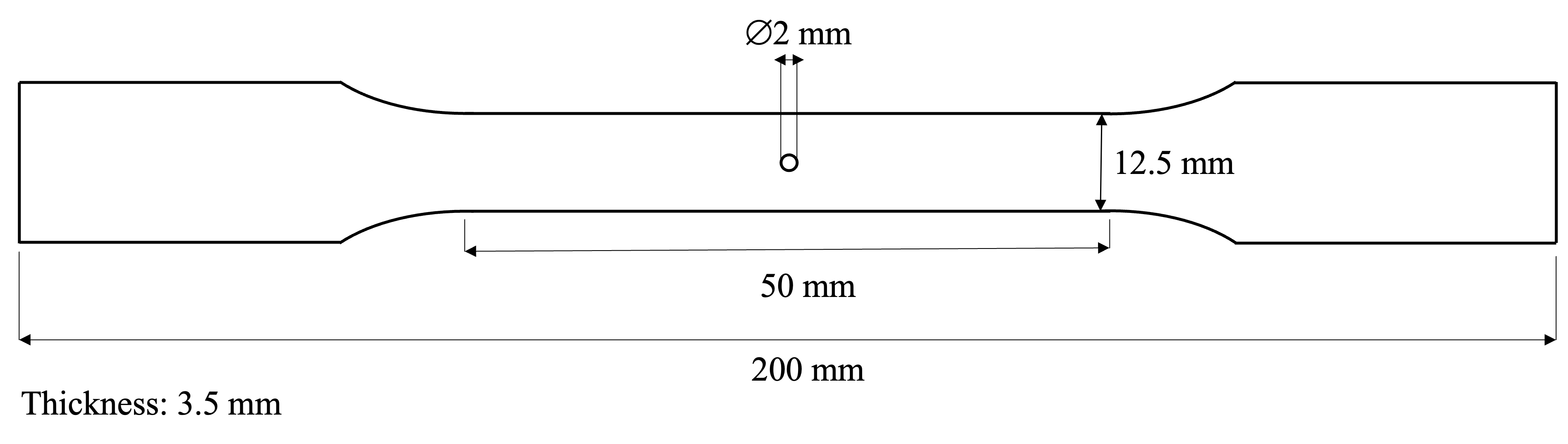}\\
    (a) \\[5pt]

    % --- second image ---
    \includegraphics[width=0.5\textwidth]{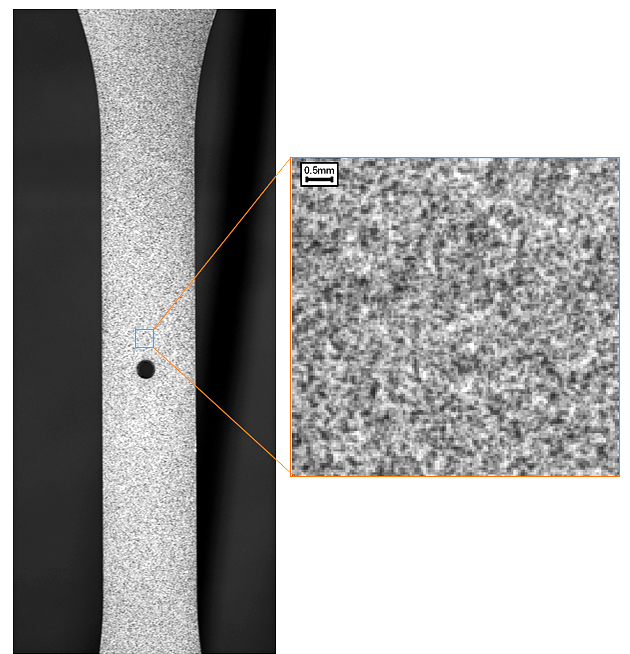}\\[2pt]
    (b) \hspace{50mm} (c)\\[5pt]

    \caption{Specimen geometry and speckle pattern: (a) Tensile specimen geometry with central hole designed for multiaxial loading and strain localization, (b) full gauge section view and (c) magnified view demonstrating speckle size distribution and contrast quality.}
    \label{fig:specimen_speckle}
\end{figure}

Specimen preparation followed a rigorous protocol to ensure dimensional accuracy and surface quality. The final specimen geometry features a central hole designed to induce multiaxial stress states and promote strain localization essential for reliable GTN parameter estimation \citep{Li2005, Dunand2012}. In accordance with ASTM E8/E8M–16a~\citep{ASTM2016}, specimens were fabricated with a gauge length of 50~mm, width of 12.5~mm, and thickness of 3.5~mm, incorporating a precisely positioned 1~mm radius hole at the center of the gauge section (Figure~\ref{fig:specimen_speckle} (a)).

The material was sourced in sheet form with a nominal thickness of 3.5~mm and stored under controlled conditions to prevent surface oxidation. Specimens were machined to final geometry exclusively by wire electrical discharge machining (EDM) using a 0.25~mm brass wire. No thermal or mechanical post-processing was applied prior to DIC surface preparation.

\subsection{Surface Preparation for DIC Analysis}
\label{subsec:surface_preparation}

Surface preparation for DIC measurements employed a systematic three-step process optimized for measurement accuracy and pattern stability. Specimens were first lightly sanded with progressively finer grit sandpaper (800, 1000, then 1200 grit) to remove machining artifacts and create a uniform surface texture conducive to paint adhesion \citep{Lecompte2006}. The sanding process was performed in a controlled manner with consistent pressure to avoid introducing surface irregularities that could compromise DIC measurement quality.

A uniform white basecoat was subsequently applied using water-soluble acrylic paint and cured for 1~hour under ambient laboratory conditions. The basecoat provides the necessary contrast foundation for the overlying speckle pattern while ensuring optimal light reflection characteristics for camera imaging \citep{Sutton2009}. Paint thickness was minimized to prevent pattern cracking during large deformations while maintaining adequate opacity for contrast optimization.

The final step involved creating a high-contrast stochastic speckle pattern using an Iwata Eclipse HP-CS airbrush operated at 2.0~bar pressure with matte black acrylic paint. The airbrush was maintained at a constant distance of 30~cm from the specimen surface, with paint diluted in a 3:1 ratio (paint:water) to achieve optimal droplet size distribution. This procedure consistently produced random patterns with speckle sizes ranging from 50 to 150~$\mu$m \citep{Pan2009}, providing optimal feature recognition for the DIC system's spatial resolution capabilities as illustrated in Figure~\ref{fig:specimen_speckle}(b) and (c). 

Pattern quality was assessed using established criteria including speckle size distribution, pattern randomness, and contrast ratio \citep{Reu2015}. The resulting patterns exhibited speckle density of approximately 15-20 speckles per correlation subset, ensuring adequate feature tracking while avoiding pattern oversaturation. All specimens were allowed to cure for 4~hours before testing to ensure pattern stability throughout the deformation process and prevent pattern degradation during mechanical loading.

The speckle pattern preparation was validated through preliminary static imaging tests to assess pattern adhesion and contrast stability under laboratory lighting conditions. Pattern quality metrics including mean gray-level gradients and subset correlation coefficients consistently exceeded recommended thresholds for reliable DIC analysis \citep{Schreier2009}. This systematic approach to surface preparation ensures consistent measurement quality across all specimens and minimizes pattern-related uncertainties in subsequent strain field calculations.

\subsection{Mechanical Testing and Data Acquisition}
\label{subsec:testing_acquisition}

Tensile tests were conducted using an Instron 5982 electromechanical, dual column test frame with 100~kN load capacity under displacement control at a constant crosshead speed of 0.02~mm/s. This loading rate corresponds to a nominal strain rate of $4 \times 10^{-4}$~s$^{-1}$ in the gauge section, ensuring quasi-static conditions appropriate for GTN model calibration while minimizing dynamic effects \citep{ASTM2016}.

F-D data were acquired using the Instron's integrated load cell and non-contact extensometer for displacement measurement, both calibrated according to ASTM standards. Calibration and replicate tests indicated a force measurement uncertainty of approximately $12~\mathrm{N}$ (about $0.1\%$ of a typical maximum force), which we adopt as the likelihood noise level ($\sigma_{\mathrm{FD}}=12~\mathrm{N}$).

\begin{figure}[htbp]
   \centering
   \includegraphics[width=0.5\textwidth]{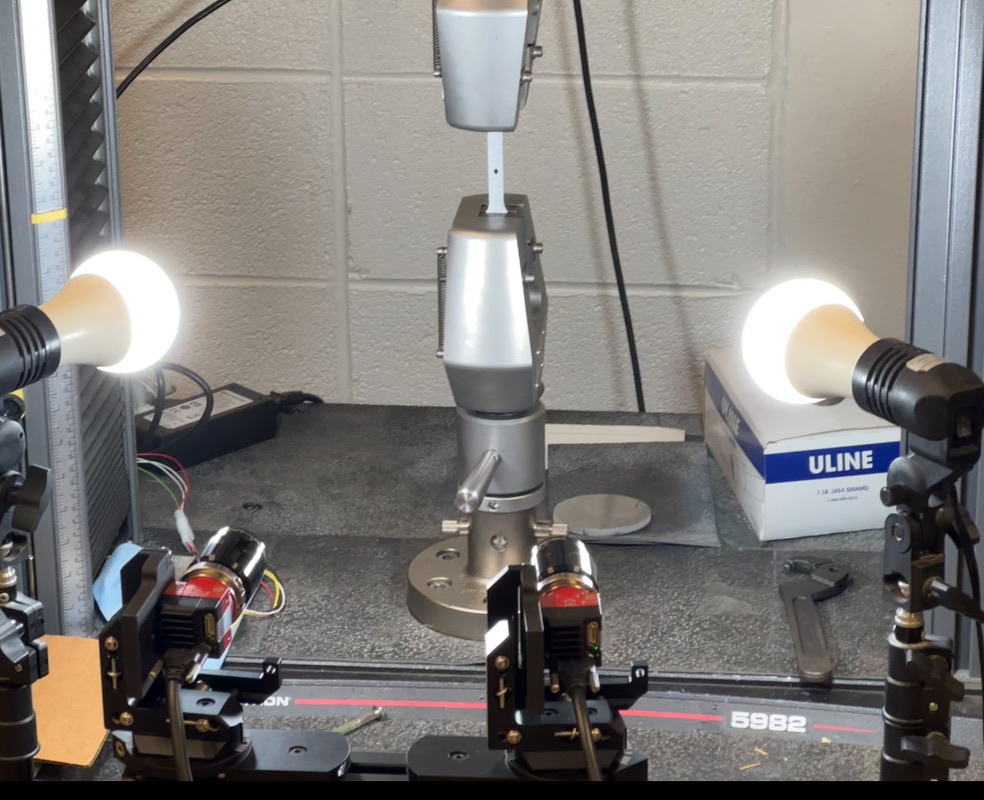}
   \caption{Stereoscopic DIC experimental setup showing dual-camera configuration with specimen positioning. The rigid mounting frame ensures stable camera positioning throughout the test duration.}
   \label{fig:dic_setup}
\end{figure}

Full-field strain measurements employed a stereoscopic DIC system comprising two Allied Vision Prosilica GT2050 cameras (2048~$\times$~1088 pixel resolution) equipped with Schneider Xenoplan 1.4/23 lenses positioned to capture three-dimensional surface deformation (Figure~\ref{fig:dic_setup}). The cameras were mounted on a rigid aluminum frame at a working distance of 400~mm from the specimen center, with an inter-camera angle of approximately 17° to enable three-dimensional deformation tracking \citep{Roux2008}.

The optical configuration provides a field of view that encompasses the entire gauge section and maintains sufficient spatial resolution for accurate strain measurement in the stress concentration region. Camera calibration was performed using a standard dot-grid calibration target and achieved calibration errors below 0.1~pixels \citep{Pan2009}. This calibration accuracy ensures reliable three-dimensional reconstruction of surface coordinates and subsequent strain calculations.

Uniform illumination was achieved using two adjustable LED panels positioned to minimize glare and shadow effects throughout the deformation process. Camera exposure times were automatically adjusted to maintain optimal image contrast as specimen geometry evolved during testing. Images were acquired at 2~Hz using VIC-Snap software and provided temporal resolution adequate for capturing damage evolution while maintaining manageable data storage requirements.

Synchronization between mechanical testing and image acquisition was achieved through electronic triggering, with both systems initiated simultaneously. Timing synchronization was verified to be within ±10~ms using external trigger signals, ensuring accurate correlation between global mechanical responses and local strain field evolution. This precise temporal coordination is important for the present study because the F-D and DIC data need to be accurately associated with each other  throughout the loading history.

\subsection{DIC Processing and Validation}
\label{subsec:dic_processing}

DIC analysis parameters were systematically optimized using VIC-3D software through parametric evaluation of subset sizes (21-61 pixels) and strain-filter sizes (9-29 pixels). The optimal configuration employed a subset size of 31~pixels and a strain-filter size of 15~pixels, providing a spatial resolution of approximately 0.17~mm while maintaining strain measurement uncertainty below 200~$\mu\varepsilon$ \citep{Lu2007}.

System noise characteristics were quantified through analysis of 20 static image sequences under identical illumination conditions. Apparent strain fluctuations in undeformed reference images indicated noise levels well below the applied testing strains, with signal-to-noise ratios exceeding 10:1 \citep{Sutton2009}. This validation ensures measured strain variations reflect actual material deformation rather than system artifacts.

Displacement vector fields were converted to Hencky (logarithmic) strain tensors to accurately represent large-deformation behavior:
\begin{equation}
\boldsymbol{\varepsilon}^{H} = \frac{1}{2}\ln\left(\mathbf{F}^T\mathbf{F}\right)
\end{equation}

\noindent where $\mathbf{F}$ is the deformation gradient tensor. This formulation provides objective strain measures appropriate for finite deformation damage analysis \citep{Belytschko2013}.

Strain data were exported as structured CSV files containing spatial coordinates, displacement components, and strain tensor components. DIC measurement uncertainty was characterized through rigid body tests, confirming displacement accuracy within ±0.002~mm and strain accuracy within the specified 200~$\mu\varepsilon$ tolerance \citep{Schreier2009}.

\subsection{Baseline Material Characterization}
\label{subsec:baseline_characterization}

Baseline constitutive material behavior of the AA6111-T4 alloy was established by conventional tensile testing of standard dog-bone specimens (i.e., without the central holes shown in Figure~\ref{fig:specimen_speckle}, but using the same overall sample geometry) \citep{ASTM2016}. These specimens were tested under identical environmental and loading conditions as the single-hole specimen used for GTN parameter estimation.

F-D data from standard tensile tests were converted to true stress-strain curves and calibrated to the Voce hardening law \citep{Voce1948}, expressed as 
\begin{equation}
\sigma = \sigma_0 + Q\left(1-\exp(-b\varepsilon^p)\right)
\label{eq:voce}
\end{equation}
where $\sigma_0$ represents the initial yield stress, $Q$ denotes the maximum hardening capacity, $b$ controls the hardening rate, and $\varepsilon^p$ is the equivalent plastic strain. This calibration was performed using the least-squares regression technique and produced the following values for the material parameters: $\sigma_0 = 165$~MPa, $Q = 136$~MPa, $b = 9.8$. These parameters capture the baseline plasticity behavior and are assumed to remain the same during damage initiation and propagation.  The plastic parameter values obtained here are in good agreement with those reported in prior literature values for AA6111-T4 \citep{Zheng2003}.

Our goal in this study is to build on this baseline plasticity model and estimate GTN model parameters for predicting damage initiation and evolution.

\section{Finite Element Implementation}
\label{sec:finite_element}

\begin{figure}[htbp]
   \centering
   \includegraphics[width=0.4\textwidth]{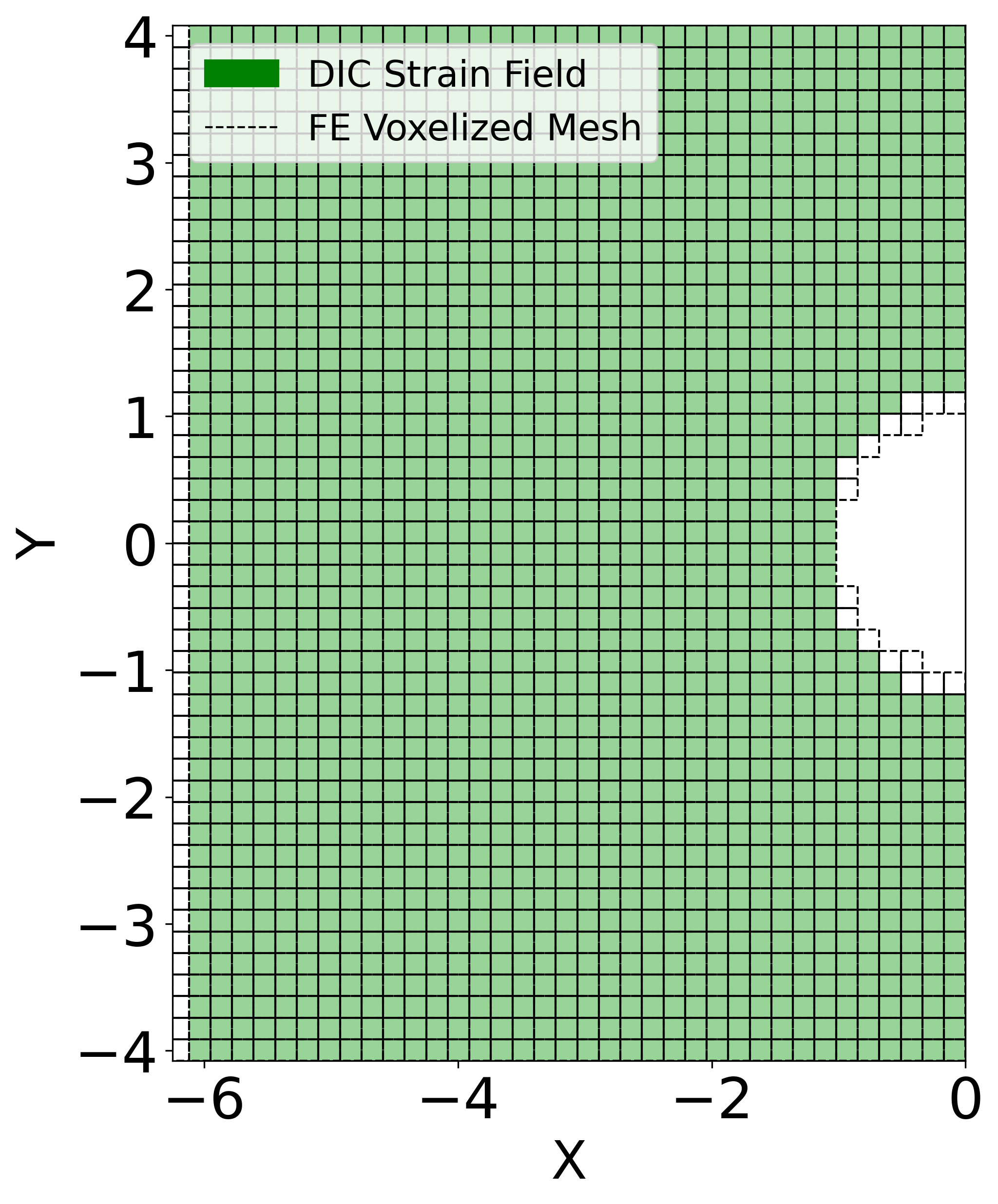} \hspace{10mm}
   \includegraphics[width=0.2\textwidth]{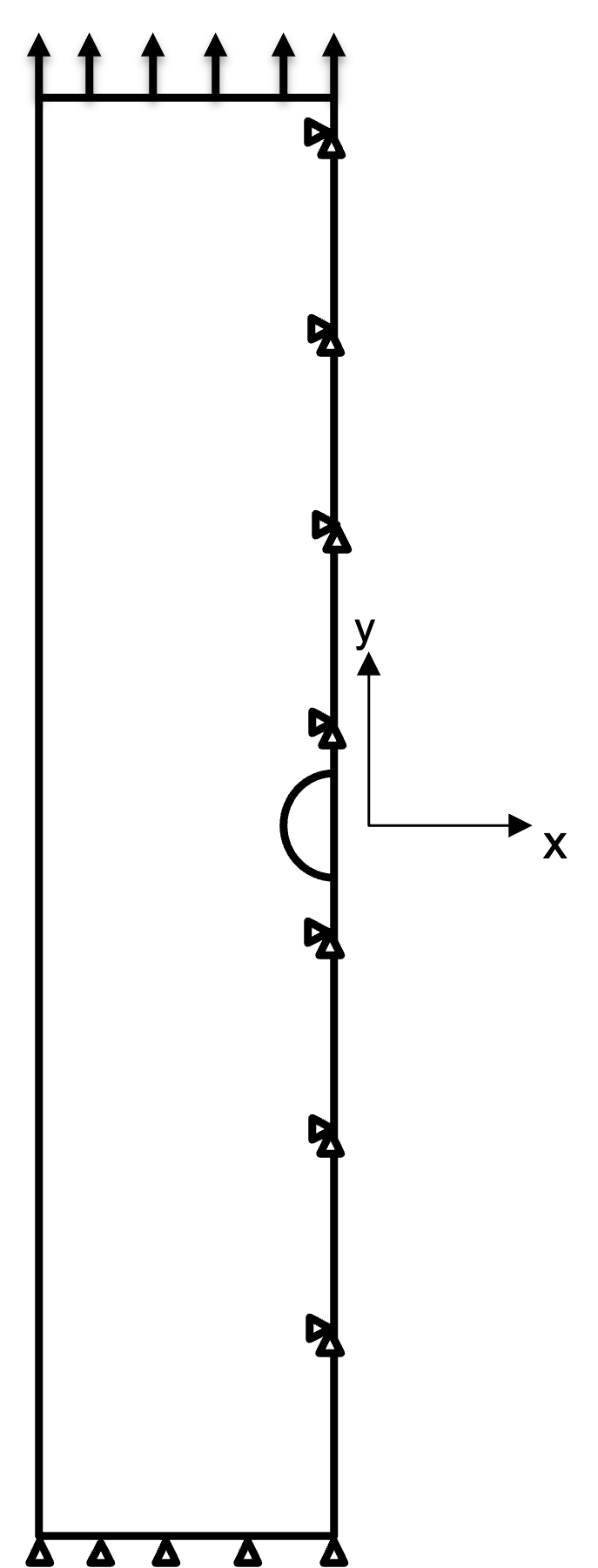}\\[2pt]
   (a) \hspace{40mm} (b) \\[2pt]
   \caption{FE model configuration: (a) Voxelized mesh showing uniform element distribution around the central hole geometry with 0.17~mm characteristic element size, and (b) boundary conditions showing the symmetry plane, vertically fixed bottom nodes, and a displacement-controlled loading condition matching the experimental setup.}
   \label{fig:fe_model}
\end{figure}

An FE model was constructed to simulate the mechanical response of the specimen featuring a central hole as shown in Figure~\ref{fig:specimen_speckle}. 
A voxelized mesh composed of uniform, 8-noded, hexahedral elements (C3D8) was employed in this study (see Figure~\ref{fig:fe_model}). The uniform  voxelized mesh architecture offers a critical advantage because the structured grid enables direct mapping between the simulation outputs and the experimental DIC measurements without the need for complex spatial interpolation schemes \citep{Sun2005, Banerjee2016, HildRoux2006}. This approach eliminates interpolation errors that commonly arise when comparing predictions from irregular (i.e., conformal) FE meshes with the DIC measurements reported on regular grids.

A characteristic element size of 0.17~mm was selected after a systematic convergence analysis was performed to determine the optimal balance between computational efficiency and solution accuracy. This element size was specifically selected to  match the DIC spatial resolution established in Section~\ref{subsec:dic_processing}. The matching resolution enables point-wise comparisons between experimental and simulated strain fields, ensuring that numerical predictions can be directly validated against experimental measurements \citep{Sun2005, Banerjee2016, Pan2009}.

Boundary and loading conditions were selected to replicate the experimental configuration as closely as possible. An $x$-symmetry constraint was applied along the central $yz$-plane with normal perpendicular to the loading direction, effectively reducing the computational domain to half the physical specimen while preserving essential mechanical behavior. A half-model configuration was selected over a quarter-model for two key reasons. First, a $y$-symmetry boundary condition would cut the hole in the region of maximum stress concentration, where the elements undergo damage and potential deletion, potentially causing boundary instability once deleted. Second, computing the entire displacement field around the hole facilitates direct validation with DIC measurements and avoids complications in determining appropriate displacement boundary conditions for a quarter model. Further boundary conditions include constraints on the bottom boundary, where the vertical ($y$) displacement is fixed, while allowing for contraction in the width and thickness ($x$ and $z$) directions. Finally, a displacement-controlled loading is applied to the top boundary equal to the experimental displacement rate of 0.02~mm/s.

The material constitutive response leverages Abaqus's porous metal plasticity implementation, providing a robust numerical framework for the GTN model presented in Section~\ref{sec:gtn_model} \citep{Abaqus2024, Erdogan2021}. This implementation offers numerically stable integration of the coupled plasticity-damage equations through adaptive time-stepping algorithms to adjust time steps during strain localization events. The material plastic behavior follows the Voce hardening law with parameters established in Section~\ref{subsec:baseline_characterization}. The GTN damage model requires specification of nine parameters, five of which were fixed based on established literature values while four others constitute the primary calibration targets for Bayesian parameter identification. The fixed parameters include the void interaction coefficients $q_1 = 1.5$, $q_2 = 1.0$, and $q_3 = q_1^2 = 2.25$ \citep{Tvergaard1984}, which govern void shape effects and coalescence behavior. These coefficients were originally calibrated by Tvergaard against detailed micromechanical FE analyses and have demonstrated robust applicability across diverse metallic materials, as evidenced by their widespread adoption in published works spanning steel grades, aluminum alloys, and other metallic materials \citep{Gholipour2019, Benzerga2004, Koplik1988}. The initial void volume fraction $f_0 = 0.001$ represents the baseline porosity level typical of wrought metallic materials in the absence of detailed microstructural characterization \citep{Garrison1987}. Fixing these well-established parameters enables focused calibration of the material-specific damage evolution parameters while maintaining computational tractability and ensuring physically meaningful parameter interactions \citep{Zhang2022, Leblond2014}. The four model parameters targeted for Sequential BI include the void nucleation strain $\varepsilon_N$, nucleation void volume fraction $f_N$, coalescence void volume fraction $f_c$, and final void volume fraction at failure $f_f$, which collectively control damage initiation, evolution, and ultimate failure behavior. The nucleation standard deviation follows the relationship $S_N = \varepsilon_N/3$, consistent with established practice in GTN parameter studies, reducing the number of independent parameter estimations while preserving realistic statistical variation in void nucleation events \citep{Gholipour2019, Chu1980, Garrison1987}.

An explicit dynamic integration scheme with controlled mass scaling is employed to enhance computational efficiency while maintaining quasi-static loading conditions \citep{Nozeres2021, Abaqus2024}. This approach proves particularly effective for capturing post-peak softening and localization phenomena characteristic of ductile failure, circumventing convergence challenges commonly encountered in implicit analyses when material instabilities develop. Mass scaling parameters are carefully selected to ensure that kinetic energy remains negligible compared to internal energy throughout the loading process, and thereby preserve the quasi-static nature of the deformation. Each simulation is terminated when the first element deletion occurs, indicating the onset of macroscopic crack initiation and defining the failure point for subsequent analysis.

During each simulation, two complementary datasets are systematically extracted to support the multimodal Sequential BI framework. Global F-D curves provide an integrated measure of macroscopic mechanical behavior throughout the deformation history, while element-averaged strains from surface elements enable direct pointwise comparison with experimental DIC measurements at corresponding displacement levels.

\section{Model Parameter Estimation}
\label{sec:parameter_estimation}
This section presents the results of GTN model parameter estimation using the sequential BI. The workflow employed in this study is illustrated in Figure~\ref{fig:methodology_simple}. The details of the analyses and computations performed in the different steps involved in this workflow are presented below.

\subsection{Surrogate Model Construction}
\label{subsec:surrogate_construction}

\subsubsection{Simulation Dataset Generation}
The construction of a comprehensive training dataset for surrogate model development necessitates systematic sampling across the GTN parameter space to ensure adequate exploration of the input domain. A sample size of $N_{train} = 400$ FE simulations was selected based on the dimensionality of the parameter space and computational resource constraints.

Latin Hypercube Sampling (LHS) was employed to generate parameter combinations across the four-dimensional space defined by the GTN damage parameters \citep{McKay1979, Santner2003}. This approach ensures uniform coverage of the parameter domain while maximizing information content from each simulation.

The GTN parameter bounds were established based on physical constraints and literature values for aluminum alloys \citep{Benzerga2004, Zhang2022}.
\begin{table}[htbp]
\centering
\caption{GTN parameter ranges for dataset generation with supporting literature.}
\label{tab:gtn_ranges}
\begin{tabular}{ll}
\hline
\textbf{Parameter} & \textbf{Range} \\
\hline
Mean nucleation strain, $\varepsilon_N$ & [0.1, 0.5] \\
Volume fraction of nucleation sites, $f_N$ & [0.01, 0.05] \\
Critical void fraction, $f_c$ & [0.01, 0.15] \\
Final void fraction, $f_f$ & [0.15, 0.35] \\
\hline
\end{tabular}
\end{table}
Each parameter combination generated through the Latin Hypercube design was subsequently employed as input to the FE model described in Section~\ref{sec:finite_element}. The simulations were executed using identical boundary conditions, mesh discretization, and loading protocols to ensure consistency across the training dataset. Displacement-controlled loading was applied at a constant rate of 0.02 mm/s to maintain quasi-static conditions throughout the deformation process.

\begin{figure}[htbp]
   \centering
   \includegraphics[width=0.5\textwidth]{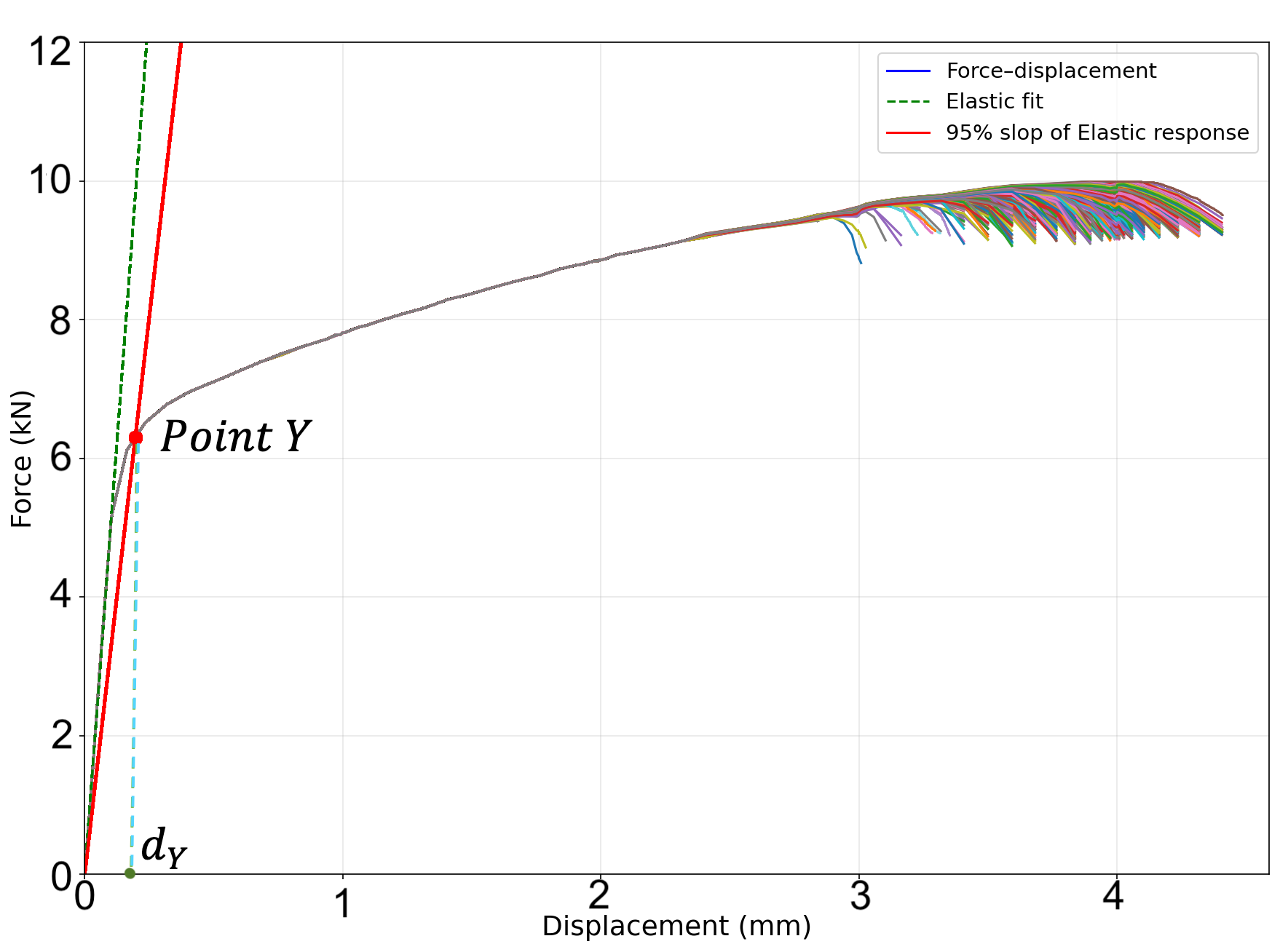}\\[2pt]
   {(a)}\\[10pt]
   \includegraphics[width=0.5\textwidth]{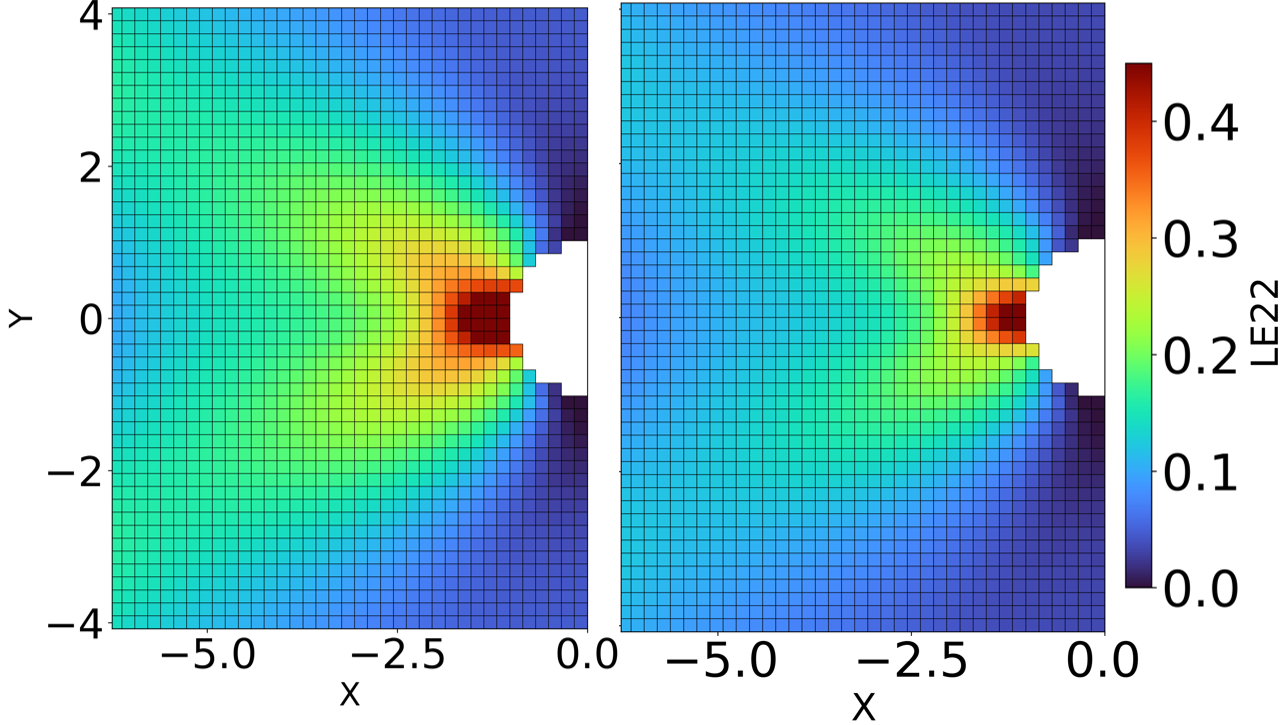}\\[2pt]
   {(b)}
   \caption{ (a) F-D curves from 400 FE simulations. Point \(Y\) is the intersection with a line from the origin having slope \(95\%\) of the initial elastic slope; the corresponding displacement \(d_Y\) is marked. (b) Strain-field examples with different levels of strain concentration.}
   \label{fig:fd}
\end{figure}

Figure~\ref{fig:fd} presents the diverse mechanical responses generated in this study for the training dataset.  
The resulting F-D curves exhibit substantial variation in peak load capacity, post-peak softening rates, and failure displacements, and confirm that the parameter sampling successfully captures a broad spectrum of damage behaviors within the specified bounds.

\subsubsection{Low-Dimensional Representations}
\label{sec:low_rep}
The F-D curves and strain fields encountered in the present application (see Figure~\ref{fig:fd}) require large feature vectors for accurate representations. For example, to represent the F-D curves shown in Figure~\ref{fig:fd} with sufficient accuracy, while including the short damage-dominant regimes that appear after the peak load, one would need high-dimensional representations. Even more specifically, if we elect to represent the F-D curve as a vector of forces at selected locations (usually on a uniform grid) on the displacement axis, a consistent and sufficiently accurate representation of the complete set of diverse F-D curves (including the damage regimes) obtained for this study would require a vector of size ~200 for each F-D curve. Similarly, each strain field (see Figure~\ref{fig:fd}(b)) obtained for this study would require an array (usually flattened into a vector while building surrogate models) with ~32,745 numbers (note that 3 different strains are stored in each voxel). Both examples above represent unwieldy high-dimensional representations for building surrogate models. Therefore, it is essential to seek suitable low-dimensional representations that offer high accuracy (i.e., minimal loss of information) at low computational cost.

Principal component analysis (PCA)~\citep{Wold1987,Jackson1991} was employed as the main tool for the low-dimensional representations needed in this work. PCA offers a versatile, low-computational cost toolset for establishing low-dimensional representations of complex signals such as F-D curves and tensorial strain fields over complex spatial domains, especially when the signals of interest are not easily represented by known analytical forms (with fittable parameters) - as in the present work. The central advantage of using PCA is that it offers an unsupervised data-driven approach to establishing a basis over complex domains that maximizes the capture of variance in the fewest number of transformed dimensions. However, the successful utilization of PCA for constructing useful low-dimensional representations is critically dependent on establishing a standardized (and discretized) representation of the diverse signals in the dataset.

Next, we describe the protocols developed and applied in this study to obtain low-dimensional representations of the F-D segments of interest required for building surrogate models. We start by recognizing that the F-D segments of interest to the present analyses start with the initiation of plasticity in the sample and end with the onset of failure. More specifically, the elastic segments preceding the start of significant plastic deformation are of no interest here,
since they exhibit identical behavior across all explored GTN parameter combinations (see Figure~\ref{fig:fd}(a)). 
While the early plastic responses also demonstrate substantial similarity across 
the training dataset, damage initiation occurred at markedly different displacement values for different combinations of GTN parameters. Therefore, the start of significant plasticity,  identified in Figure~\ref{fig:fd}(a) as Point Y, was selected as the standardized starting point for isolating the F-D segments for our analysis. This specific point was identified as the intersection of a linear response through the origin with a slope equal to 95\% of the slope in the initial elastic response with the predicted F-D curve (see Figure~\ref{fig:fd}(a)). The end point for the segment of interest in the F-D curve was selected as the end point from the FE simulation, which was defined in Section~\ref{sec:finite_element} to correspond to the first instance of element deletion during the simulation.

One of the main challenges in applying PCA to the relevant F-D segments arises because the segments span different ranges on the displacement axis, due to variations in failure points seen in Figure~\ref{fig:fd}(a)). This challenge was addressed in this work using the following steps: (i) Linearly map the displacement range $(d_Y, d_f)$ for each segment of interest to (0,1), where $d_Y$ is the displacement at Point Y and $d_f$ is the displacement at failure in the FE predicted curve. (ii) Find force values at 200 equally spaced displacement values over the selected range, and represent them as a vector $\{F\}$. Note that the set $(d_Y, d_f, \{F\})$ provide a discretized representation of each F-D segment of interest; this set can adequately reconstruct the original F-D segment within a small discretization error. (iii) Transform the individual features of the $\{F\}$ vector to z-scores, producing a vector $\{Z\}$; this constitutes a one-to-one mapping between $\{F\}$ and $\{Z\}$ vectors. This transformation ensures that each force component receives equal consideration in the subsequent PCA step. (iv) Perform PCA on the complete training set of 400 $\{Z\}$ vectors. Note that this transformation also constitutes a one-to-one map. (v) Truncate the PCA output, denoted as PC score vector \{$\alpha$\} for each F-D segment, to a consistent truncation level selected to capture the desired level of variance in the dataset. In the present work, it was observed that 7 PC scores (i.e., the first 7 values in each \{$\alpha$\} vector) preserved over 99\% of the variance in the training set. This represents an effective dimensionality reduction from 200 to 8 (7 PC scores and $d_f$).

PCA was also performed on the strain fields of interest obtained from the 400 FE simulations to establish the low-dimensional representations needed for the downstream surrogate modeling effort. Because the same FE mesh was utilized for all 400 FE simulations (see Figure~\ref{fig:fe_model}) and element-averaged strain components were extracted near the sample surface, the strain field descriptions were already standardized for PCA. The remaining decision to be made was regarding which snapshot(s) of the evolving strain field should be utilized for the analyses.

\begin{figure}[htbp]
   \centering
   \includegraphics[width=0.5\textwidth]{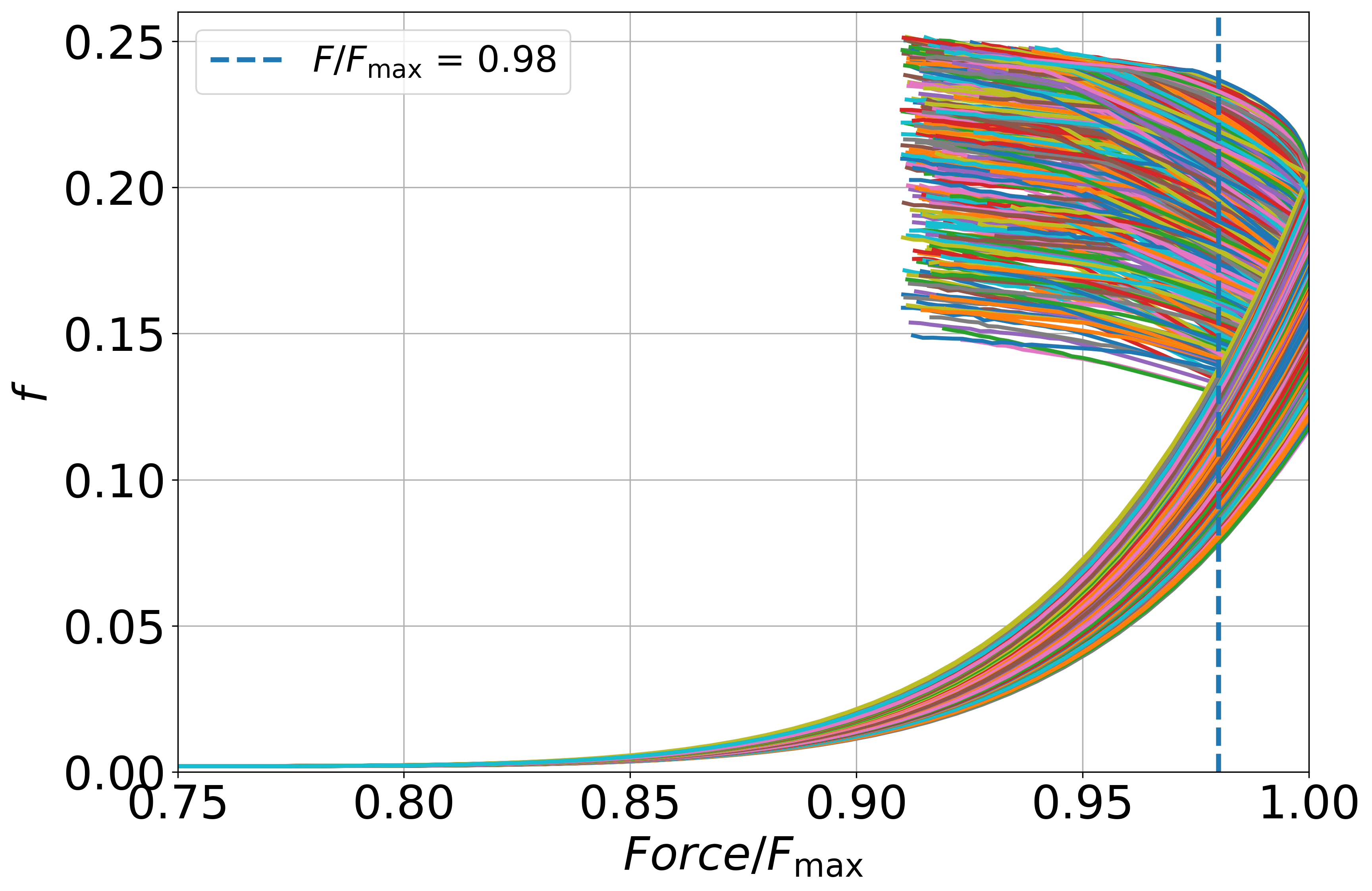}
   \caption{Maximum void volume fraction evolution across all 400 FE simulations. Curves show consistent behavior below $F/F_{\max} = 0.98$, with numerical oscillations appearing in some simulations beyond this threshold.}
   \label{fig:vvf}
\end{figure}

From the training set of 400 FE simulations, it was observed that the failure load during post-peak softening was in the range of 90-94\% of the peak force (the mean ratio was 92.2\% and the median was 92.1\%). Our goal was to extract strain fields well before fracture, but after the peak load. This is to ensure that we are capturing adequate details of damage in the sample, while not being adversely affected by potential numerical artifacts arising from the fracture process. Within this narrow damage regime (between the peak load and fracture), damage evolution characteristics were examined across the entire training set by tracking the evolution of the highest value of the void volume fraction ($f$) among all elements in the mesh, which is the main internal damage variable in the GTN model. Figure~\ref{fig:vvf} presents the evolution of $f$  for all 400 simulations versus $F/F_{\max}$, where $F$ is the current applied force and $F_{\max}$ is the peak force achieved during each simulation. From this figure, it was observed that numerical oscillations begin to appear in some simulations after $F/F_{\max} = 0.98$, indicating the onset of computational artifacts that could compromise strain field reliability. Below this threshold, the curves exhibit consistent, smooth evolution across the entire simulation dataset. Therefore, $F/F_{\max} = 0.98$ was selected as the optimal extraction point for this analyses. Since the reliable damage regime in the FE simulations is relatively short (only a 2\% drop from the peak load), it was decided to utilize only one strain field for the analyses performed in this study. It should be noted that the methods developed and demonstrated here are not limited by this selection; i.e., they are already generalized to accommodate multiple strain fields, as might be needed for future studies. 

For the FE mesh used in this study, the strain field of interest includes the in-plane strain tensor components $(\varepsilon_{11}, \varepsilon_{12}, \varepsilon_{22})$ from 10{,}915 surface elements. These strain values are concatenated into vectors $\mathbf{s}\in\mathbb{R}^{32745}$, and subjected to PCA. Scaling protocols were applied to ensure that no single strain component dominates the variance structure during the PCA dimensionality reduction \citep{Generale2021}. This is important to preserve the essential spatial damage patterns while maintaining balanced contributions from all strain components in the reduced-order representation. The scaling factors required to balance the variance contributions in our dataset were established to be 1.87 for $\varepsilon_{11}$ and 2.79 for $\varepsilon_{12}$. The first five principal components explained $\ge 99\%$ of the variance while preserving essential spatial damage patterns for surrogate modeling. The PC score vector used in this study for each strain field is therefore denoted as a vector $\{\beta\}$ with 5 PC scores, effectively resulting in a dimensionality reduction from 32745 to 5.

\subsubsection{Gaussian Process Regression}

The first step (Step 1 in both sequential updates described in Figure 1) requires the construction of surrogate models from the FE-produced training set. The inputs for these surrogate models are the GTN parameters (denoted by $\boldsymbol{\theta}$) and the outputs are the respective PC scores of the F-D curves and the DIC strain fields. Single-output Gaussian Process (SOGP) regression models were constructed for each PC score involved in these models. Although it is possible to build multi-output GP models for the present application, such approaches were found to significantly increase the computational cost involved without a commensurate benefit in the accuracy of the overall model. For each PC score (each element of $\{\alpha\}$ and $\{\beta\}$), a GP model was established of the form (following Eq. (7)):
\begin{equation}
f(\boldsymbol{\theta}) \sim \text{GP}(0, k(\boldsymbol{\theta}, \boldsymbol{\theta}'))
\label{eq:sogp}
\end{equation}
where the use of zero mean functions is justified by the fact that the PC scores are naturally zero-centered.

The covariance structure employed in this work is a radial basis function kernel with automatic relevance determination (ARD):
\begin{equation}
k(\boldsymbol{\theta}, \boldsymbol{\theta}') = \sigma_f^2 \exp\left(-\frac{1}{2}\sum_{i=1}^{4}\frac{(\theta_i - \theta_i')^2}{\ell_i^2}\right) + \sigma_n^2\delta(\boldsymbol{\theta}, \boldsymbol{\theta}')
\end{equation}
\noindent where $\sigma_f^2$ controls the overall magnitude variation for each output, while the individual length scales $\ell_i$ capture the sensitivity of each output variable to the respective input variable. Small values of $\ell_i$ suggest rapid function variation and high sensitivity, while large values indicate low sensitivity. The noise variance $\sigma_n^2$ accounts for uncertainties implicit in the training set.

Hyperparameter optimization (i.e., obtaining optimized values of $\sigma_f^2$, $\ell_i$, and $\sigma_n^2$ for each output) was carried out with a Latin-hypercube design of 15 initialization points~\citep{McKay1979} with the L-BFGS-B algorithm~\citep{Zhu1997} to maximize the log-likelihood while avoiding convergence to local optima. Bounds were selected based on preliminary sensitivity analysis and established GP practices: length scales $[10^{-3}, 10^2]$ to span parameter ranges, signal variance $[10^{-6}, 10^2]$ to capture response magnitudes, and noise variance $[10^{-8}, 10^{-1}]$ to reflect FE simulation precision. The evaluation of model performance relied on prediction accuracy over the test dataset, using a 75\%–25\% train–test split.

The surrogate model performance for the F-D segments of interest was assessed through measures of mean absolute error (MAE) and normalized mean absolute error (NMAE) for each F-D curve in the test set, calculated as
\begin{equation}
   \text{MAE} = \frac{1}{N}\sum_{i=1}^{N}|F_i - \hat{F}_i| 
\end{equation}

\begin{equation}
    \text{NMAE} = \frac{1}{N}\sum_{i=1}^{N}\frac{|F_i - \hat{F}_i|}{F_{\text{average}}} \times 100
\end{equation}
\noindent where $F_i$ represents the true force values from FE simulations, $\hat{F}_i$ denotes the corresponding surrogate model predictions (reconstructed from the predicted set of 7 PC scores), and $N = 200$ denotes the number of discretized points in each F-D curve, and $F_{\text{average}}$ denotes the ensemble average of the ground-truth values from the training set.

\begin{figure}[htbp]
    \centering
    % First row with two images side by side - this is figure (a)
    \includegraphics[width=0.4\textwidth]{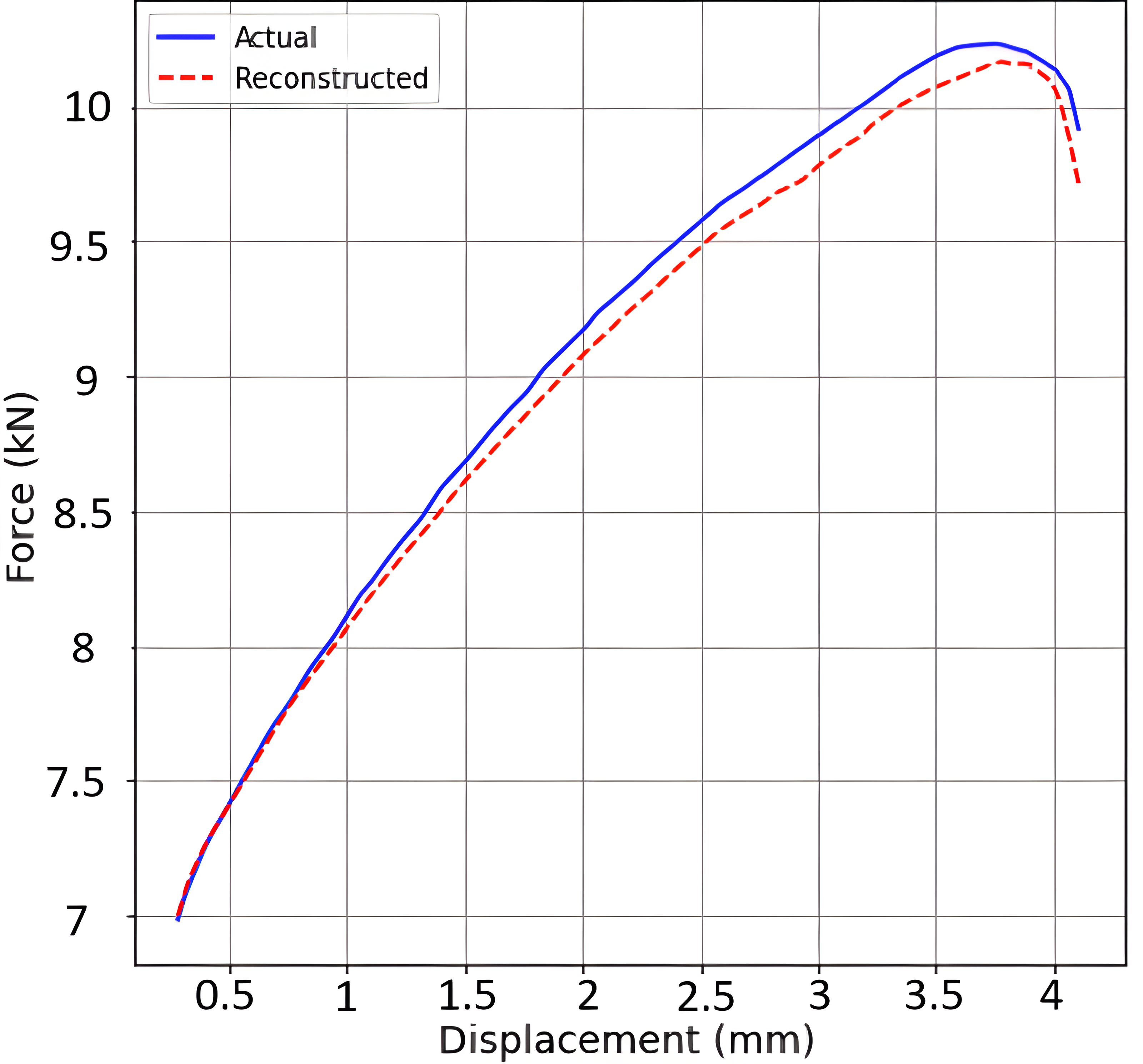}
    \includegraphics[width=0.4\textwidth]{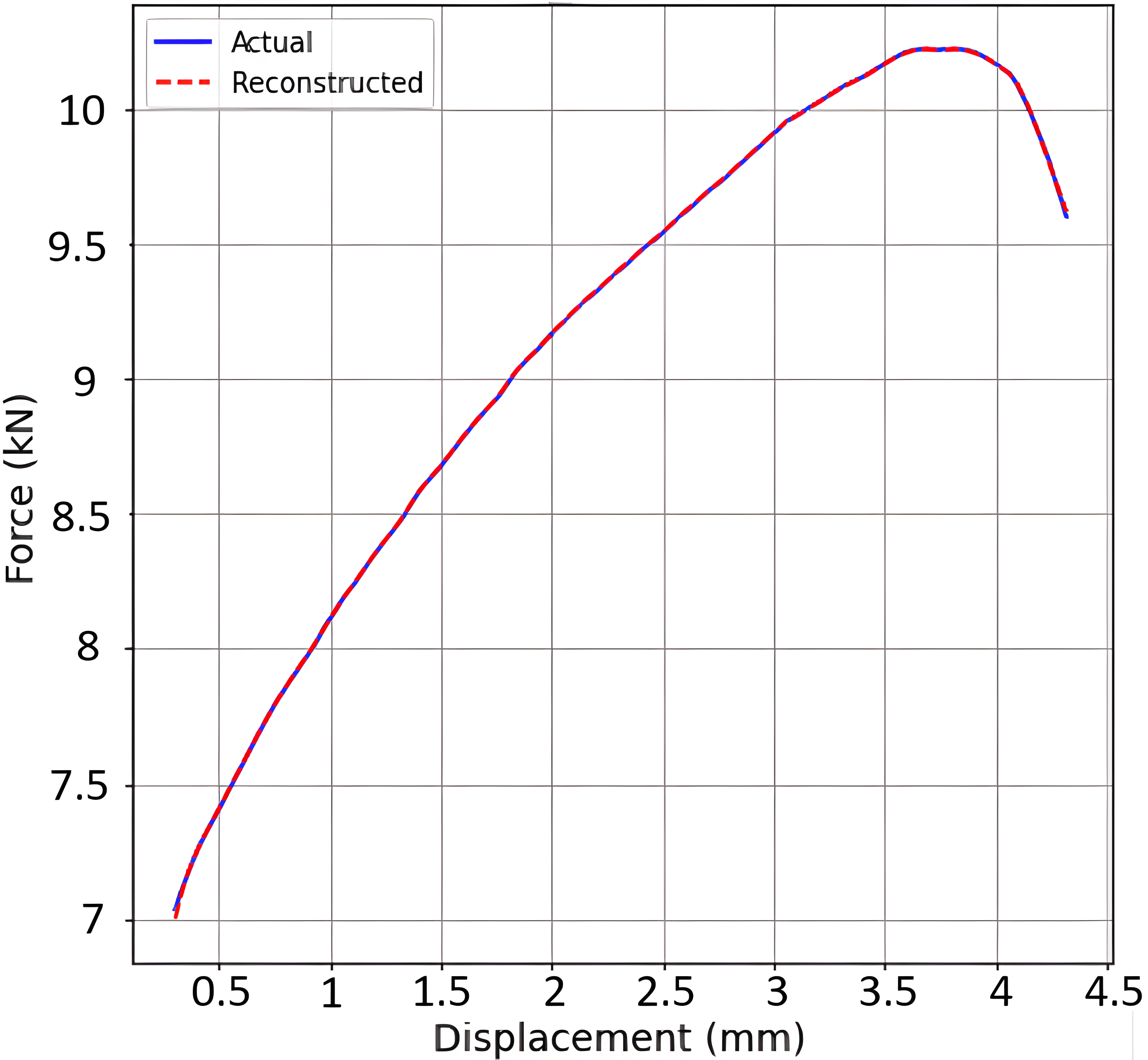}
    
    % Add some vertical space between rows
    \vspace{0.1cm}
    (a)\\[10pt]
    % Second row with one centered image - this is figure (b)
    \includegraphics[width=0.5\textwidth]{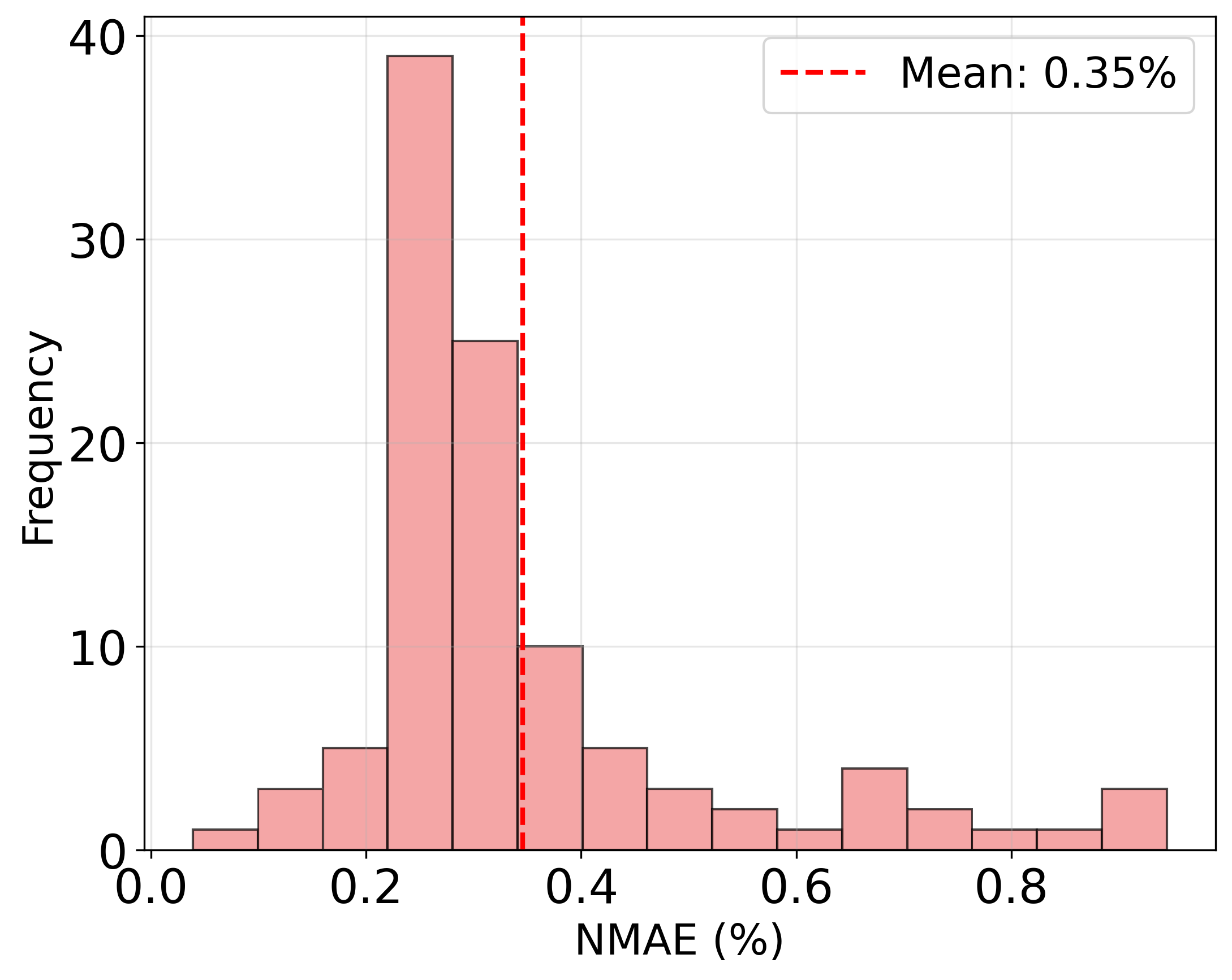}\\[2pt]
    (b)\\[10pt]
    % Second row with one centered image - this is figure (b)
    %\includegraphics[width=0.5\textwidth]{pf_nmae.png}\\[2pt]
    %(c)\\[10pt]
    \caption{Surrogate model accuracy showing (a) Left: best-case and right: worst-case prediction accuracy (NMAE = 0.07\% and 0.85\% respectively), demonstrating robust emulation across the parameter space, and (b) the distribution of NMAE error over the test set}

    \label{fig:surrogate_validation}
\end{figure}

Figure~\ref{fig:surrogate_validation} summarizes the predictive accuracy of the F-D surrogate model on the test set. Figure~\ref{fig:surrogate_validation}(a)
 compares surrogate reconstructions (dashed red) and FE ground-truth (solid blue) for two representative \emph{unseen} test cases - the best case (NMAE = \(0.07\%\)) and the worst case (NMAE = \(0.85\%\)) predictions - demonstrating close agreement, from yield through the softening regime. NMAE distribution in Figure~\ref{fig:surrogate_validation}(b) reveals high accuracy of the surrogate model over the entire test set: the mean NMAE is \(0.35\%\), the 95th percentile is \(<0.8\%\), and the maximum is \(<1\%\). These plots confirm the high fidelity of the surrogate models produced to predict the F-D curves. It should be noted that the use of PCA made it possible to build these relatively light high-fidelity models (only six hyperparameters for each of the seven PC scores representing the F-D curves).

Surrogate models were similarly constructed for the PC scores of the strain fields. Their accuracy was assessed using component-wise error metrics. For each test point (strain field) $m$ and each in-plane strain component $\varepsilon_{ij}$ ($i,j\in\{1,2\}$), we report a single per-field error by averaging over spatial locations $p=1,\dots,P$ in the region of interest:
\begin{equation}
\mathrm{MAE}^{(m)}_{\varepsilon_{ij}}
= \frac{1}{P}\sum_{p=1}^{P}
\left|\varepsilon^{(m)}_{ij,p}-\hat{\varepsilon}^{(m)}_{ij,p}\right|,
\end{equation}
\begin{equation}
\mathrm{NMAE}^{(m)}_{\varepsilon_{ij}}
= \frac{100}{\varepsilon_{\mathrm{ref},ij}}\,\mathrm{MAE}^{(m)}_{\varepsilon_{ij}}.
\end{equation}
\noindent Here $\varepsilon^{(m)}_{ij,p}$ denotes the FE ground-truth value for strain component $ij$, at spatial location $p$, within test case $m$. Similarly, $\hat{\varepsilon}^{(m)}_{ij,p}$ represents the corresponding GP surrogate prediction for the same component, location, and test case. The normalization constant $\varepsilon_{\mathrm{ref},ij}$ is computed from the training data as the average absolute magnitude of the corresponding in-plane strain component,

\begin{equation} 
\varepsilon_{\mathrm{ref},ij} = \frac{1}{N_{\mathrm{train}}}\sum_{q=1}^{N_{\mathrm{train}}}\left|\varepsilon_{ij}\right|. 
\end{equation}

\begin{figure}[htbp]
    \centering
    \includegraphics[width=\textwidth]{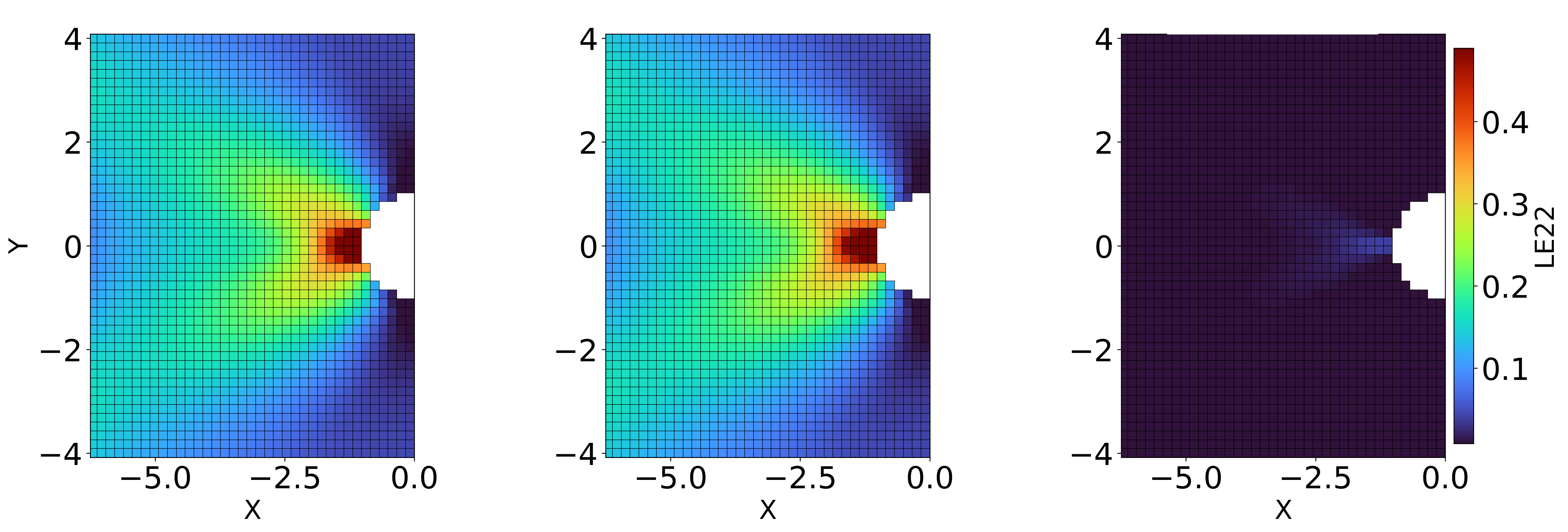}\\[2pt]
    (a)\\[2pt]
    \includegraphics[width=\textwidth]{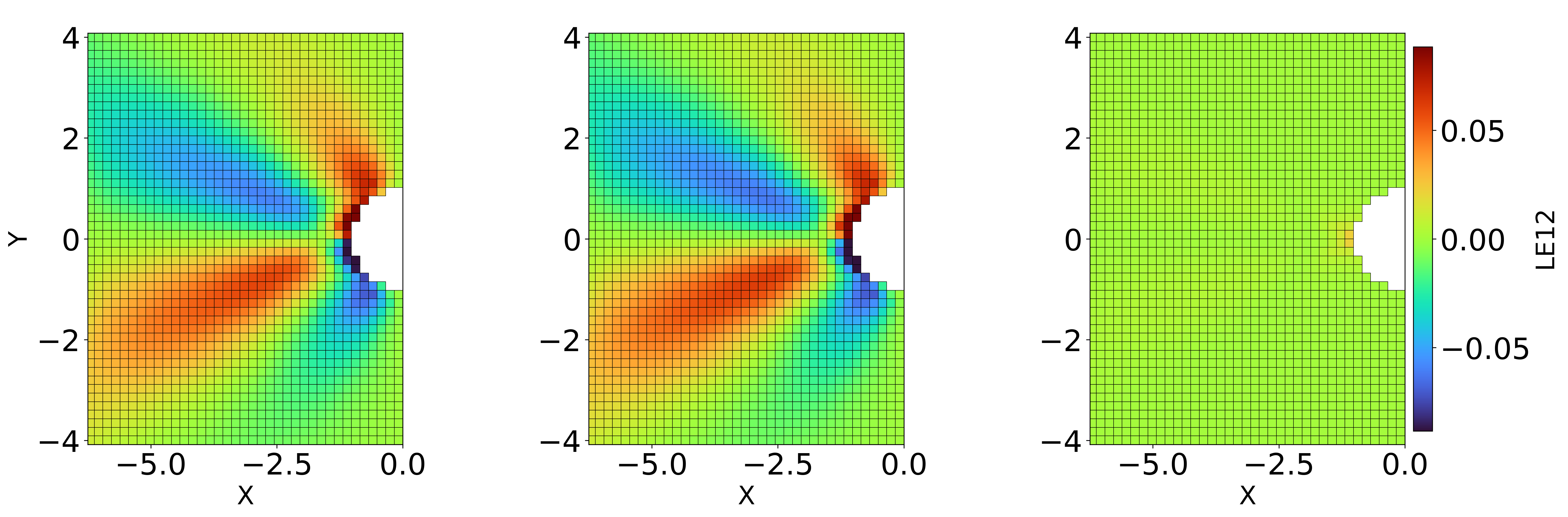}\\[2pt]
    (b)\\[2pt]
    \includegraphics[width=\textwidth]{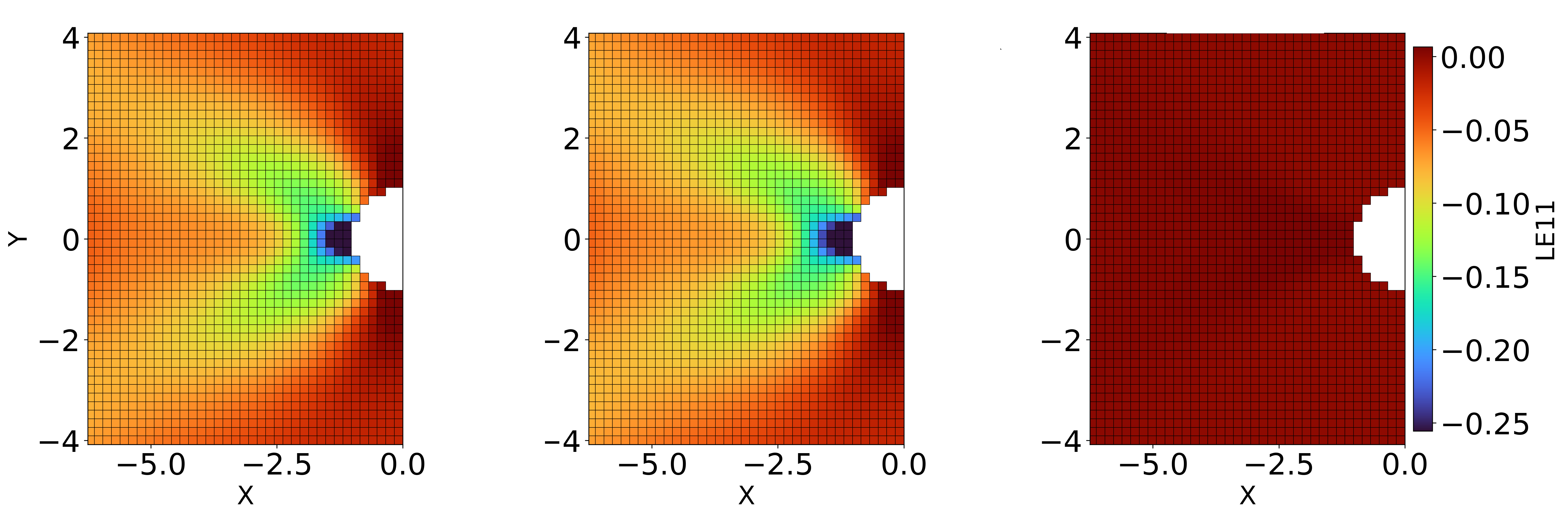}\\[2pt]
    (c)\\[2pt]
    
    \caption{Worst-case spatial validation for the three in-plane strain components. Columns show (left) FE strain, (center) GP surrogate prediction, and (right) absolute difference for each component. Rows: (a) $\varepsilon_{22}$, (b) $\varepsilon_{12}$, (c) $\varepsilon_{11}$. Maximum component errors (NMAE): $\varepsilon_{22}=2.8\%$, $\varepsilon_{12}=3.5\%$, $\varepsilon_{11}=2.6\%$.}

    \label{fig:strain_component_validation}
\end{figure}

The strain field surrogate models demonstrate the power of principal component analysis for efficient emulation of complex spatial responses. By reducing full strain fields to just 5 PC scores, the framework enables accurate Gaussian process surrogates using only hyperparameters per output (i.e., per PC score). This is remarkable keeping in mind that the predicted strain field has a total of 32,745 strain quantities. Component-wise NMAE values (see Table \ref{tab:strain_componentaccuracy}) remain below 1.5\% for all strain components (0.867\% for $\varepsilon_{11}$, 1.422\% for $\varepsilon_{12}$, 0.855\% for $\varepsilon_{22}$), with corresponding MAE values on the order of $10^{-4}$–$10^{-3}$. Most importantly, even in worst-case scenarios where component-wise NMAE reaches 2.6-3.5\%, the PCA-based surrogate accurately preserves the essential localization patterns around the central hole—precisely the spatial damage features that govern GTN parameter sensitivity (Figure~\ref{fig:strain_component_validation}). This combination of computational efficiency and pattern fidelity makes the approach practical for  BI while maintaining the physical insight necessary for reliable parameter identification.

\begin{table}[ht]
\centering
\caption{Strain field surrogate model accuracy by component.}
\label{tab:strain_componentaccuracy}
\begin{tabular}{lcc}
\toprule
Component & MAE (Mean) & NMAE (\%) (Mean) \\
\midrule
$\varepsilon_{11}$ & 0.000628 & 0.867 \\
$\varepsilon_{12}$ & 0.000421 & 1.422 \\
$\varepsilon_{22}$ & 0.001249 & 0.855 \\
\midrule
Overall            & 0.000766 & 1.048 \\
\bottomrule
\end{tabular}
\end{table}

\newpage

\subsection{Sequential Bayesian Inference Results}
\label{subsec:calibration_results}

This section reports the sequential BI results. We first summarize the details of  T-MCMC sampling implementation along with diagnostics. We then present the two single-dataset estimated posteriors (one incorporating experimental F-D only; the other experimental DIC only) to quantify the information provided by each modality. Finally, we compare two estimated posteriors that incorporate experimental data but differ in their assimilation order (see Figure 1) and quantify the sequence effects. As validation of the posteriors obtained in this work, we input the estimated parameter values into the FE model and compare the simulated and experimental F–D and strain-field responses.

\subsubsection{T-MCMC sampling implementation}
We start by establishing uniform priors for the GTN parameters, $\boldsymbol{\theta}=\{\varepsilon_N,f_N,f_c,f_f\}$  within the bounds shown in Table~\ref{tab:gtn_ranges}. These bounds were established based on previous studies~\citep{Benzerga2004,Zhang2022}. We enforce a physically motivated constraint, $(f_c<f_f)$, by setting the prior density to zero whenever $(f_c\ge f_f)$.

Next, we transform the experimental responses (F–D and DIC) into their PC scores using the PC bases already established in Section~\ref{sec:low_rep}. These will be denoted as $\mathbf{s}^{\mathrm{FD}}=[\alpha_1,\ldots,\alpha_7,d_f]$ and $\mathbf{s}^{\mathrm{DIC}}=[\beta_1,\ldots,\beta_5]$, respectively. A candidate $\boldsymbol{\theta}$ is then evaluated by comparing the experimentally observed PC scores to the GP-predicted PC scores corresponding to $\boldsymbol{\theta}$. The details of the Gaussian log-likelihood computation and the treatment of measurement noise in PC-score space are provided in~\ref{app:likelihood}.

Transitional MCMC was employed with the annealing parameter $\gamma$ increased from 0 to 1; at each intermediate level, a few short Metropolis–Hastings updates were performed to sample from the tempered target distribution~\citep{Hastings1970}. This annealed sequence stabilizes mixing and improves exploration of the posterior relative to a single-stage random-walk sampler. We ran 8 independent runs (chains) for between-run diagnostics. Convergence was assessed with the split Gelman-Rubin statistic $\hat{R}$ and the effective sample size (ESS) \citep{Gelman2013}. All runs achieved split-$\hat{R}<1.05$ (most $<1.02$), and the minimum ESS per parameter exceeded 6500. For additional details of the T-MCMC implementation, the reader is referred to prior publications \citep{Ching2007,Venkatraman2025}.

Maximum a posteriori (MAP) estimates and 95\% highest posterior density (HPD) intervals~\citep{Gelman2013} are extracted from the sampled posteriors and reported in this work. The MAP represents the most probable parameter value (the mode of the posterior distribution), while the 95\% HPD provides the narrowest interval containing 95\% of the posterior probability mass for that parameter (the uncertainty band).

\subsubsection{Single-Dataset Posteriors}
\label{subsubsec:update1}
\begin{figure}[htbp]
    \centering
    \begin{subfigure}[b]{0.5\textwidth}
        \centering
        \includegraphics[width=\textwidth]{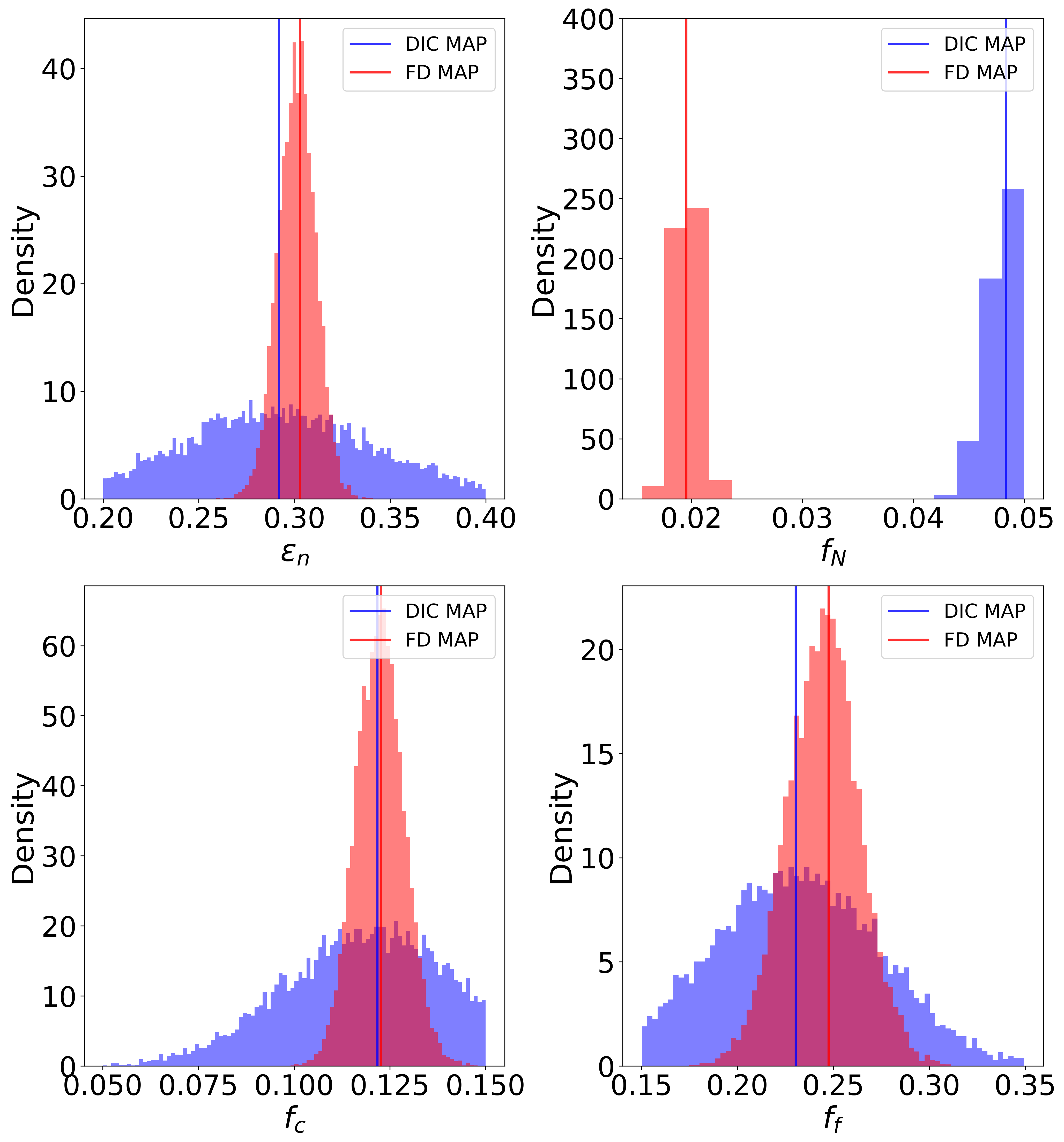}
        \caption{}
    \end{subfigure}
    \vspace{1em}
    \begin{subfigure}[b]{0.5\textwidth}
        \centering
        \includegraphics[width=0.9\textwidth]{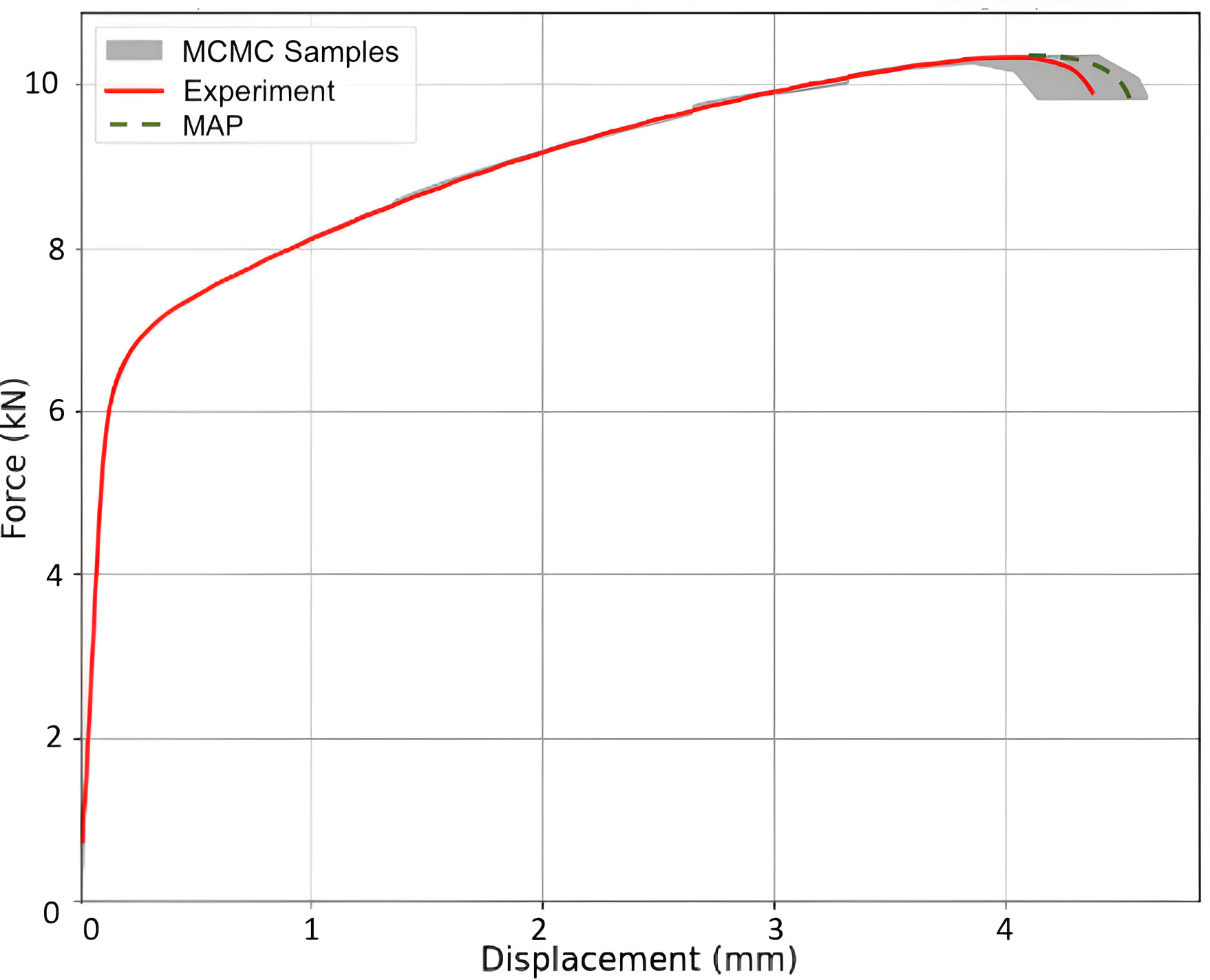}
        \caption{}
    \end{subfigure}
    \caption{(a) Single-dataset posterior distribution with MAP (vertical lines): DIC (blue) vs F–D (red). (b) F–D prediction from the F–D-only posterior: shaded band (MCMC samples), the experimental curve (red), and MAP parameter combination (dashed line). Values for MAP and HPDs are also summarized in Table~\ref{tab:single_stats}.}
    \label{fig:single}
\end{figure}

With the prior and likelihood specified, two single-modality T-MCMC updates (F-D only and DIC only) were run, both satisfying the convergence criteria described above. The resulting posteriors are unimodal (Figure~\ref{fig:single}a), with the F–D only posterior consistently more concentrated; corner plots show no cross-parameter correlation, meaning each parameter is identifiable on its own—tightening one does not force changes in the others—simplifying interpretation ( see\ref{app:corner_plots}, Figs.~\ref{fig:corner_fd}–\ref{fig:corner_dicfd}).
 The DIC-only posterior shows an upward shift in the nucleation fraction: MAP $f_N=0.0484$ versus $0.0195$ for F-D—and the $95\%$ HPDs do not overlap: DIC $[0.0453,\,0.0500]$ versus F-D $[0.0177,\,0.0217]$. For $\varepsilon_N$, the MAPs are close ($0.2919$ versus $0.3030$) and the $95\%$ HPDs overlap; for $f_c$ and $f_f$, the posterior means are similar but the DIC-only intervals are visibly broader. In the DIC-only run, several posteriors reach the bounds of the uniform priors (Figure~\ref{fig:single}a; Table~\ref{tab:gtn_ranges}). This reflects prior support rather than a sampling artifact and suggests limited identifiability (prior-dominated inference) under DIC-only data.

Using posterior samples from $p(\boldsymbol{\theta}\mid\mathbf{y}^{\mathrm{FD}})$, the GP surrogate reconstructs the experimental F-D curve across yield, hardening, and the post-peak region (Figure~\ref{fig:single}b), confirming that F-D alone strongly constrains the global response. The experimental curve lies within the MCMC envelope (posterior predictive band). F-D only run provides a sharper constraint (narrower HPDs, accurate F-D reconstruction) than DIC only run, which has larger uncertainty bands. This motivates assimilating F-D first, followed by DIC.

\begin{table}[htbp]
\centering
\small
\caption{Single-dataset posteriors (MAP and 95\% HPD).}
\label{tab:single_stats}
\begin{tabular}{lcccc}
\toprule
 & \multicolumn{2}{c}{F–D only} & \multicolumn{2}{c}{DIC only} \\
\cmidrule(lr){2-3}\cmidrule(lr){4-5}
Parameter & MAP & 95\% HPD & MAP & 95\% HPD \\
\midrule
$\varepsilon_N$ & 0.3030 & [0.2821, 0.3203] & 0.2919 & [0.2047, 0.3753] \\
$f_N$           & 0.0195 & [0.0177, 0.0217] & 0.0484 & [0.0453, 0.0500] \\
$f_c$           & 0.1227 & [0.1104, 0.1348] & 0.1217 & [0.0820, 0.1500] \\
$f_f$           & 0.2476 & [0.2075, 0.2819] & 0.2304 & [0.1540, 0.3047] \\
\bottomrule
\end{tabular}
\end{table}

\subsubsection{Order Sensitivity}
\label{subsubsec:update2}

\begin{figure}[htbp]
    \centering
    \includegraphics[width=0.6\textwidth]{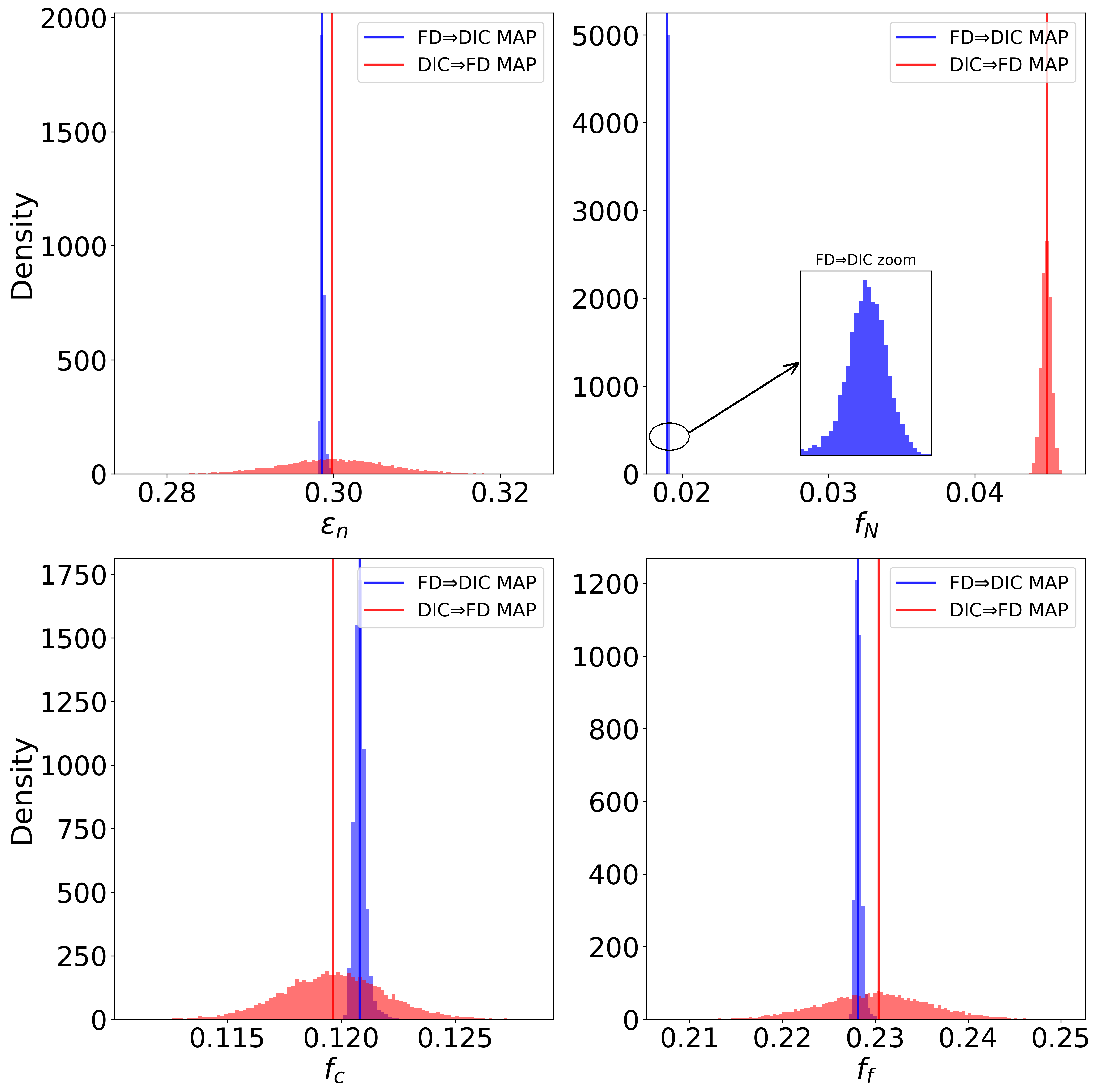}
    \caption{Final posteriors for GTN parameters under two orders--F-D$\rightarrow$DIC (blue) vs DIC$\rightarrow$F-D (red). Vertical lines: MAP; inset highlights the extremely tight $f_N$ band for F-D$\rightarrow$DIC.}
    \label{fig:final}
\end{figure}

In order to use the single dataset posteriors (from Update 1 in Figure~\ref{fig:single}) as the prior for the sequential update (i.e., Update 2 in Figure~\ref{fig:final}) with the other dataset while preserving bounds and non-Gaussian shape, we reparameterize each bounded parameter $\theta_i\in(a_i,b_i)$ via the logit map

\begin{equation}
z_i=\log\!\frac{\theta_i-a_i}{\,b_i-\theta_i\,},\quad i=1,\dots,4,
\label{eq:logit}
\end{equation}
\noindent A Gaussian kernel density estimate is then fit to the transformed samples, and the Update 2 prior is defined via a change of variables ($\boldsymbol{\theta}$) (Jacobian of \eqref{eq:logit}) \citep{Silverman1986,Scott2015}. Upon mapping the KDE prior back to \(\boldsymbol{\theta}\) via the inverse–logit, the same truncation is applied to preserve the constraint \(f_c<f_f\), i.e., support is restricted to \(\{\boldsymbol{\theta}: f_c<f_f\}\).

\begin{figure}[htbp]
    \centering
    \begin{subfigure}[b]{0.5\textwidth}
        \centering
        \includegraphics[width=\textwidth]{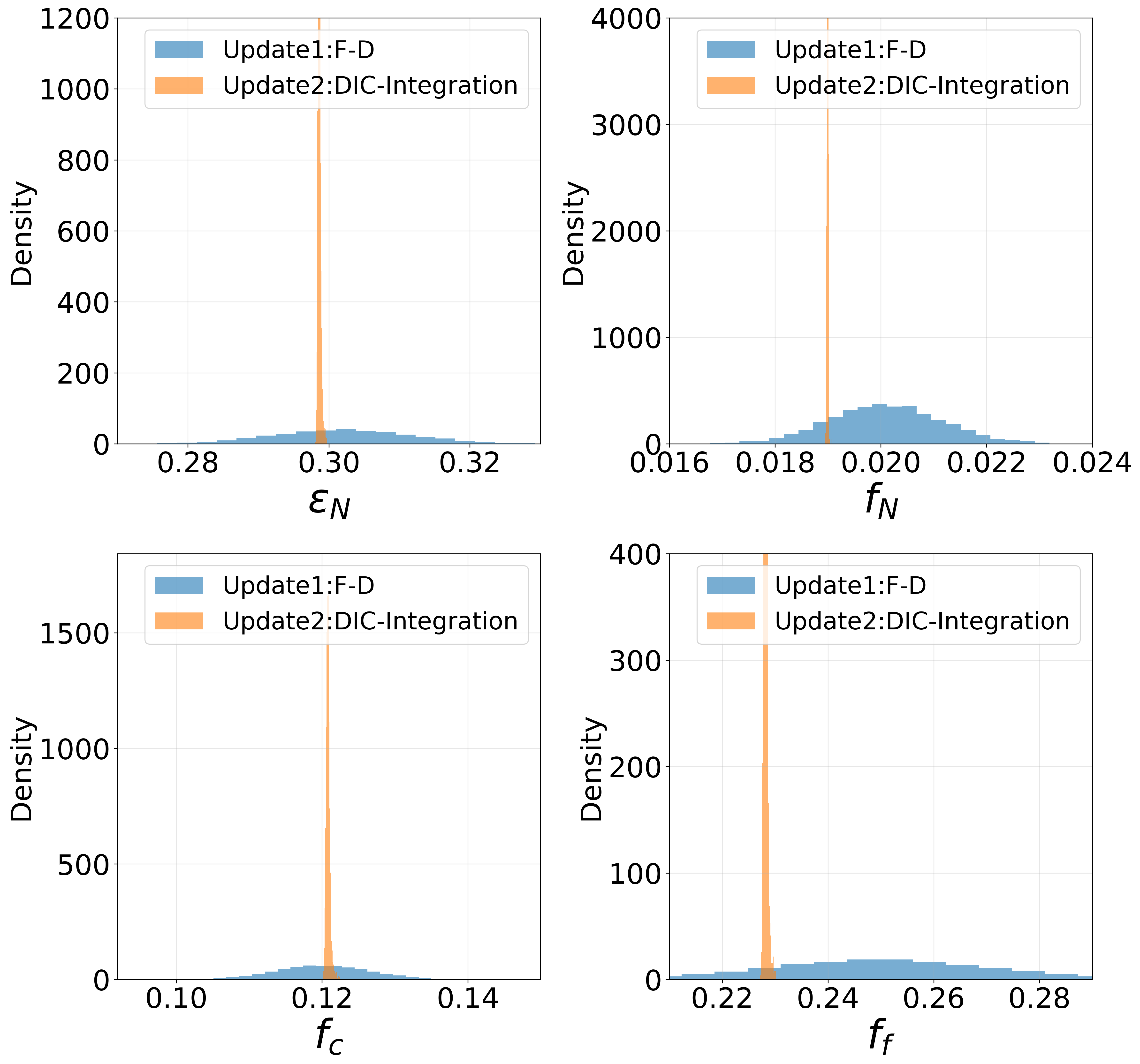}
        \caption{}
    \end{subfigure}
    \vspace{1em}
    \begin{subfigure}[b]{0.5\textwidth}
        \centering
        \includegraphics[width=\textwidth]{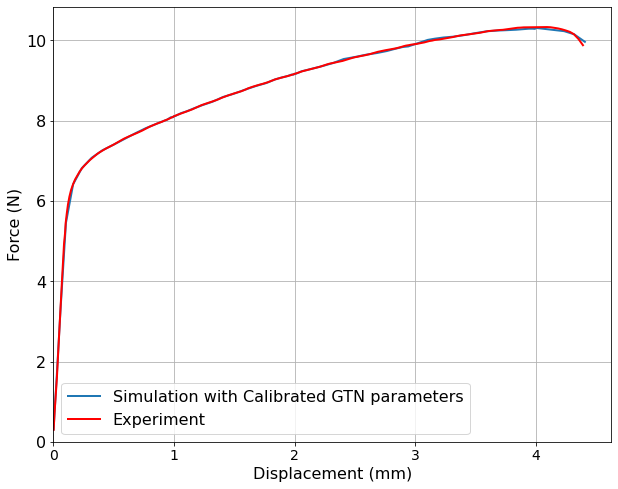}
        \caption{}
    \end{subfigure}
    \caption{(a) Posterior densities before (blue) and after the DIC update (orange). (b) F-D predicted with the final MAP versus experiment.}
    \label{fig:seq_update}
\end{figure}

The following results evaluate whether assimilation is order-invariant or shows order-sensitive identifiability. Table~\ref{tab:sequence_comparison} and Figure~\ref{fig:final} compare the final posteriors under the two update orders. F–D$\rightarrow$DIC sequence yields tight 95\% HPDs with negligible MAP shifts relative to Update~1, whereas the DIC$\rightarrow$F–D sequence leaves substantially broader marginals (order-of-magnitude increases for $\varepsilon_N$ and $f_N$, and near order-of-magnitude for $f_c$ and $f_f$). Overlaying Update~1 (F–D only) with the final F–D$\rightarrow$DIC posteriors (Figure~\ref{fig:seq_update}a) shows that the DIC step drives the tightening: compared to F–D only, the 95\% HPD widths shrink by $\sim$1–2 orders of magnitude, with the strongest contraction for $f_N$. MAP shifts are summarized in Table~\ref{tab:update_fd_then_dic}, indicating that DIC primarily constrains nucleation/coalescence parameters—lowering $f_f$ and tightly concentrating $f_N$—while $\varepsilon_N$ and $f_c$ change little.

This pattern reflects the complementary information in the two modalities. The global F-D curve provides stabilizing constraints across all parameters, whereas the post-peak DIC field is highly sensitive to nucleation and coalescence localization. When the sequence begins from the DIC-only posterior (Table~\ref{tab:single_stats}, Figure~\ref{fig:single}a), the subsequent F-D step lacks sufficient leverage to substantially narrow the broad intervals. Conversely, starting from F-D produces a compact posterior which is then sharpened by the DIC step.

\begin{table}[htbp]
\centering
\small
\caption{Sequential posteriors (MAP and 95\% HPD) for the two assimilation orders.}
\label{tab:sequence_comparison}
\begin{tabular}{lcccc}
\toprule
 & \multicolumn{2}{c}{F-D$\rightarrow$DIC} & \multicolumn{2}{c}{DIC$\rightarrow$F-D} \\
\cmidrule(lr){2-3}\cmidrule(lr){4-5}
Parameter & MAP & 95\% HPD & MAP & 95\% HPD \\
\midrule
$\varepsilon_N$ & 0.2986 & [0.2983, 0.2990] & 0.2997 & [0.2879, 0.3131] \\
$f_N$           & 0.01899 & [0.0189, 0.0190] & 0.0449 & [0.0442, 0.0455] \\
$f_c$           & 0.1209 & [0.1203, 0.1213] & 0.1196 & [0.1153, 0.1242] \\
$f_f$           & 0.2281 & [0.2275, 0.2288] & 0.2304 & [0.2182, 0.2410] \\
\bottomrule
\end{tabular}
\end{table}

\begin{table}[htbp]
\centering
\small
\caption{Sequential update starting with F–D: MAP and 95\% HPD after Update~1 (F–D only) and after Update~2 (F–D$\rightarrow$DIC).}
\label{tab:update_fd_then_dic}
\begin{tabular}{lcccc}
\toprule
 & \multicolumn{2}{c}{Update 1: F--D only} & \multicolumn{2}{c}{Update 2: F--D$\rightarrow$DIC} \\
\cmidrule(lr){2-3}\cmidrule(lr){4-5}
Parameter & MAP & 95\% HPD & MAP & 95\% HPD \\
\midrule
$\varepsilon_N$ & 0.3030 & [0.2821, 0.3203] & 0.2986 & [0.2983, 0.2990] \\
$f_N$           & 0.0195 & [0.0177, 0.0217] & 0.01899 & [0.0189, 0.0190] \\
$f_c$           & 0.1227 & [0.1104, 0.1348] & 0.1209 & [0.1203, 0.1213] \\
$f_f$           & 0.2476 & [0.2075, 0.2819] & 0.2281 & [0.2275, 0.2288] \\
\bottomrule
\end{tabular}
\end{table}

\begin{figure}[htbp]
    \centering

    \includegraphics[width=\textwidth]{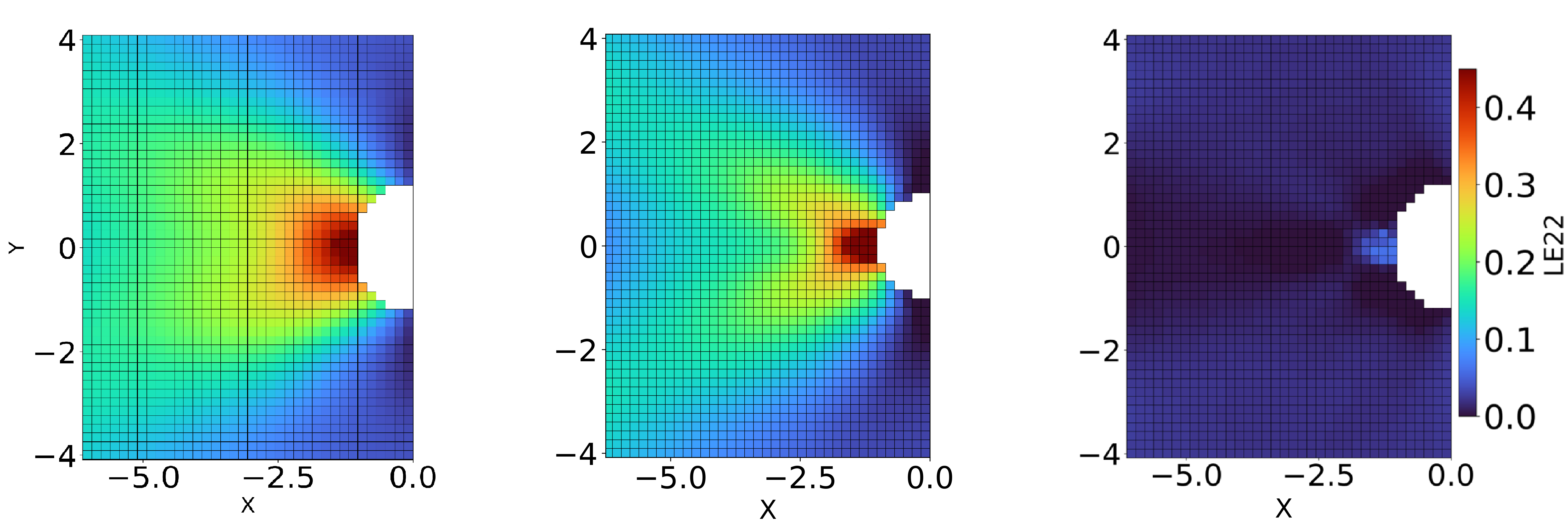}\\[2pt]
    (a)\\[2pt]
    \includegraphics[width=\textwidth]{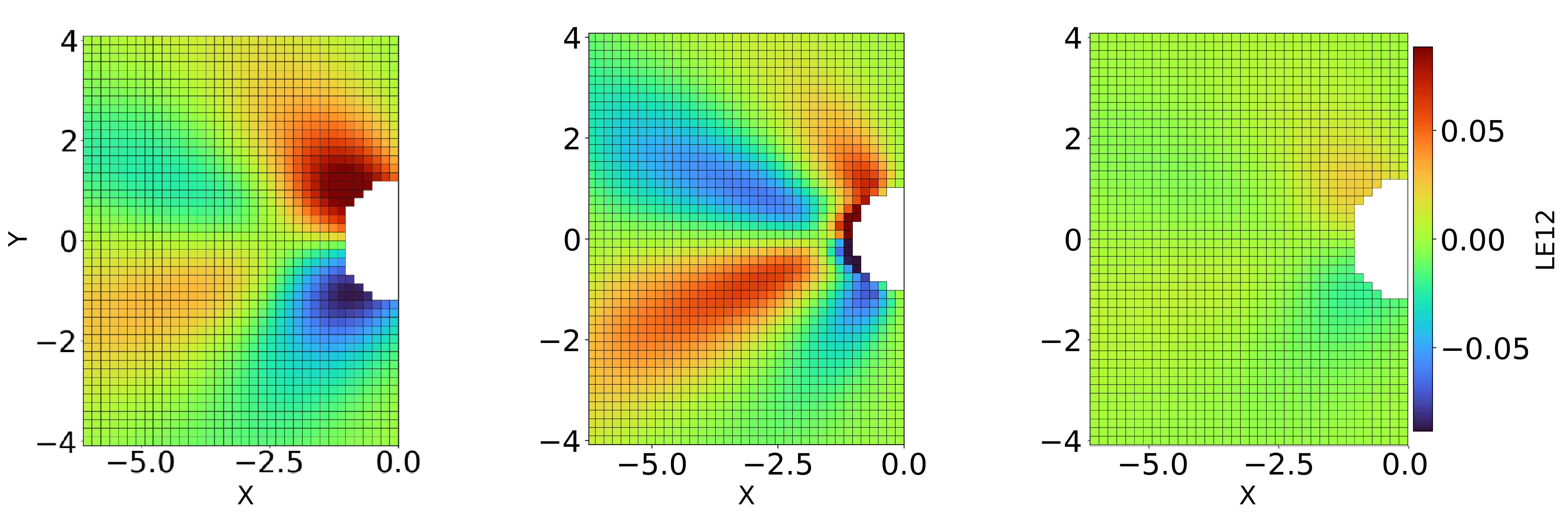}\\[2pt]
    (b)\\[2pt]
    \includegraphics[width=\textwidth]{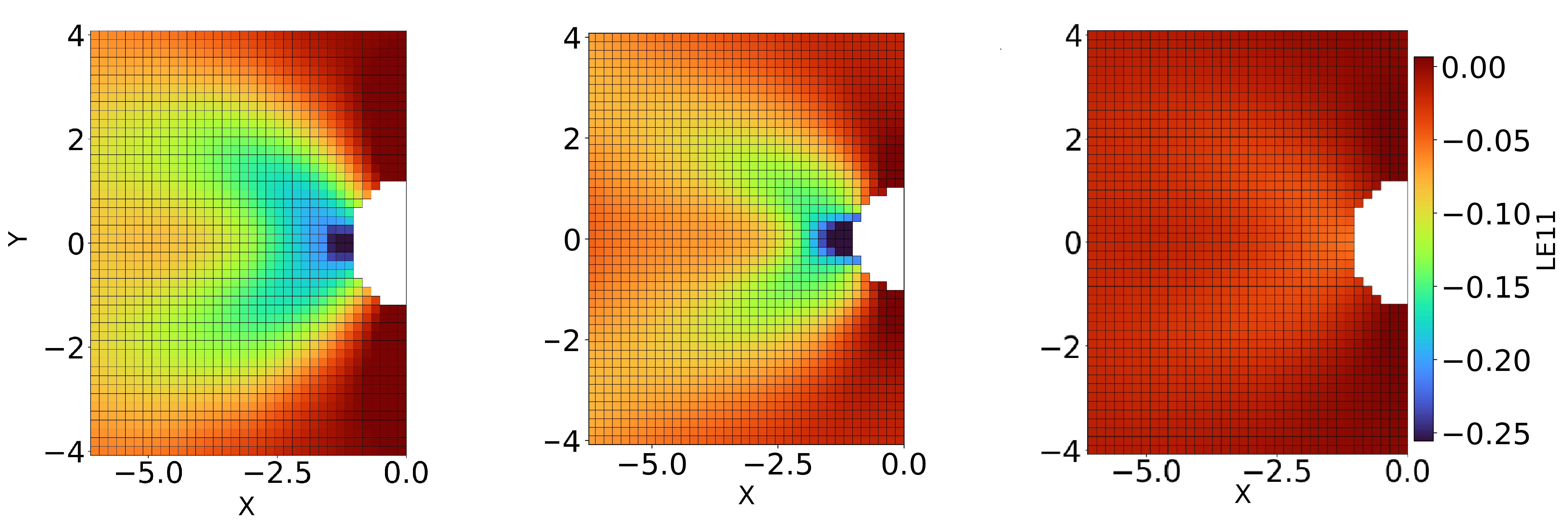}\\[2pt]
    (c)\\[2pt] 
    
    \caption{Spatial validation at $F=0.98\,F_{\max}$ (post-peak snapshot). Columns show (left) experimental DIC strain, (center) estimated model prediction, and (right) FE-DIC difference for each component. (a): $\varepsilon_{22}$; (b): $\varepsilon_{12}$; (c): $\varepsilon_{11}$. Over the region of interest: $\mathrm{MAE}(\varepsilon_{22})=8.91\times10^{-3}$, $\mathrm{MAE}(\varepsilon_{12})=1.50\times10^{-2}$, $\mathrm{MAE}(\varepsilon_{11})= 9.54 \times10^{-2}$.}
    \label{fig:dic_val}
\end{figure}

It is seen that the estimated parameters indeed accurately reproduce the measured F–D curve across yield, hardening, and the brief post-peak softening window (Figure~\ref{fig:seq_update}(b)), with NMAE of only 0.3\%. Note the remarkable improvement in the match between the experiment and the model in the post-peak F-D curve, when compared to the MAP values obtained from a calibration with only the F-D data in Figure~\ref{fig:single}(b). Additionally, they also reproduce the strain-fields  at $F/F_{\max}=0.98$ quite accurately (Figure~\ref{fig:dic_val}). Specifically, the peak magnitudes and the locus of localization around the hole are well captured, and the difference maps exhibit small, spatially confined residuals consistent with DIC noise levels and mesh resolution. Over the region of interest, the mean absolute errors are $\mathrm{MAE}(\varepsilon_{22})=8.91\times10^{-3}$, $\mathrm{MAE}(\varepsilon_{12})=1.50\times10^{-2}$, and $\mathrm{MAE}(\varepsilon_{11})=9.54\times10^{-2}$. These results confirm that the  posterior obtained by the proposed Bayesian framework not only reduces parameter uncertainty but also produces simulations that match both global (i.e., F-D) and local (i.e., DIC) observables.

\begin{figure}[htbp]
    \centering
    \includegraphics[width=0.46\textwidth]{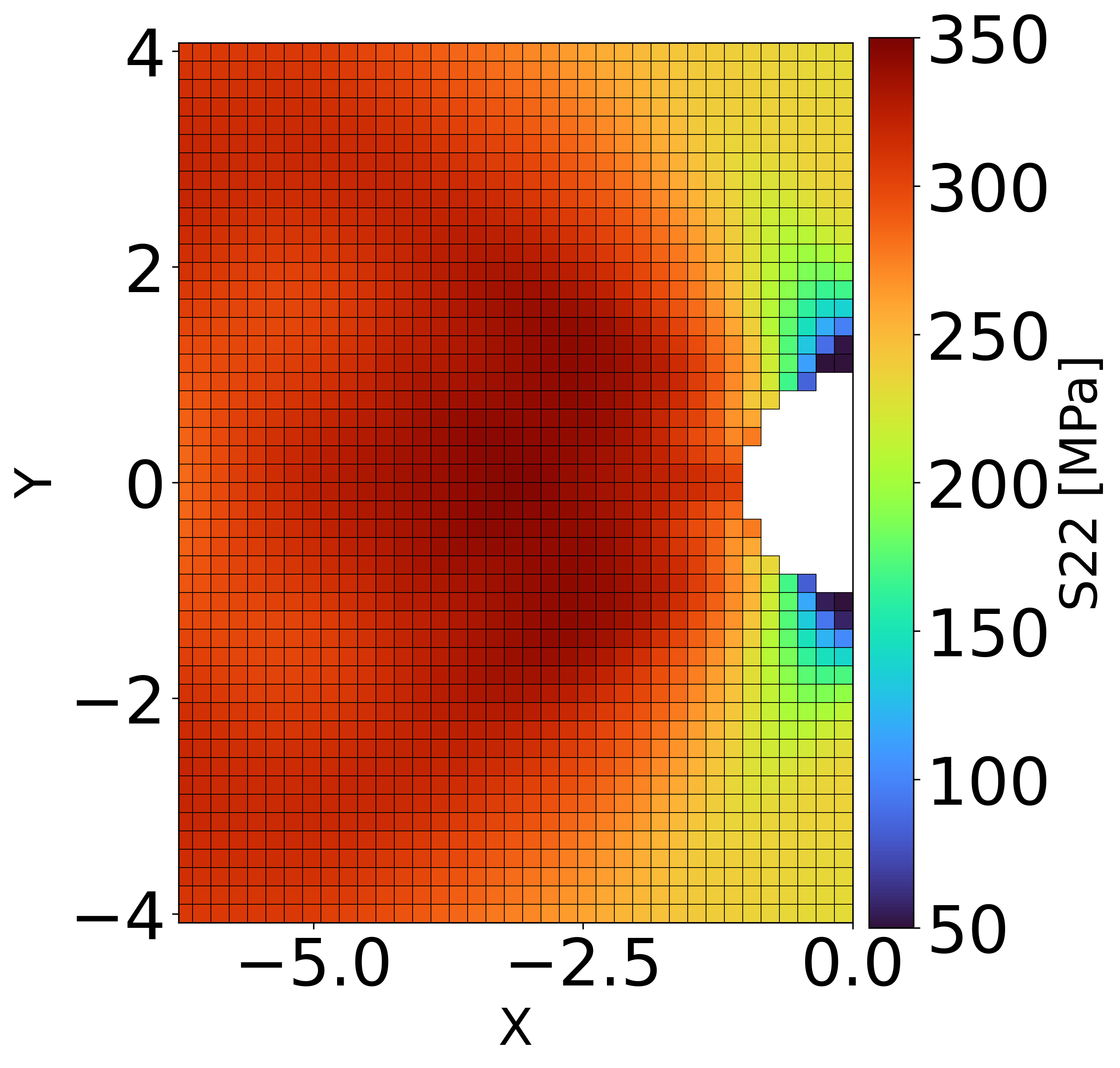}\hfill
    \includegraphics[width=0.46\textwidth]{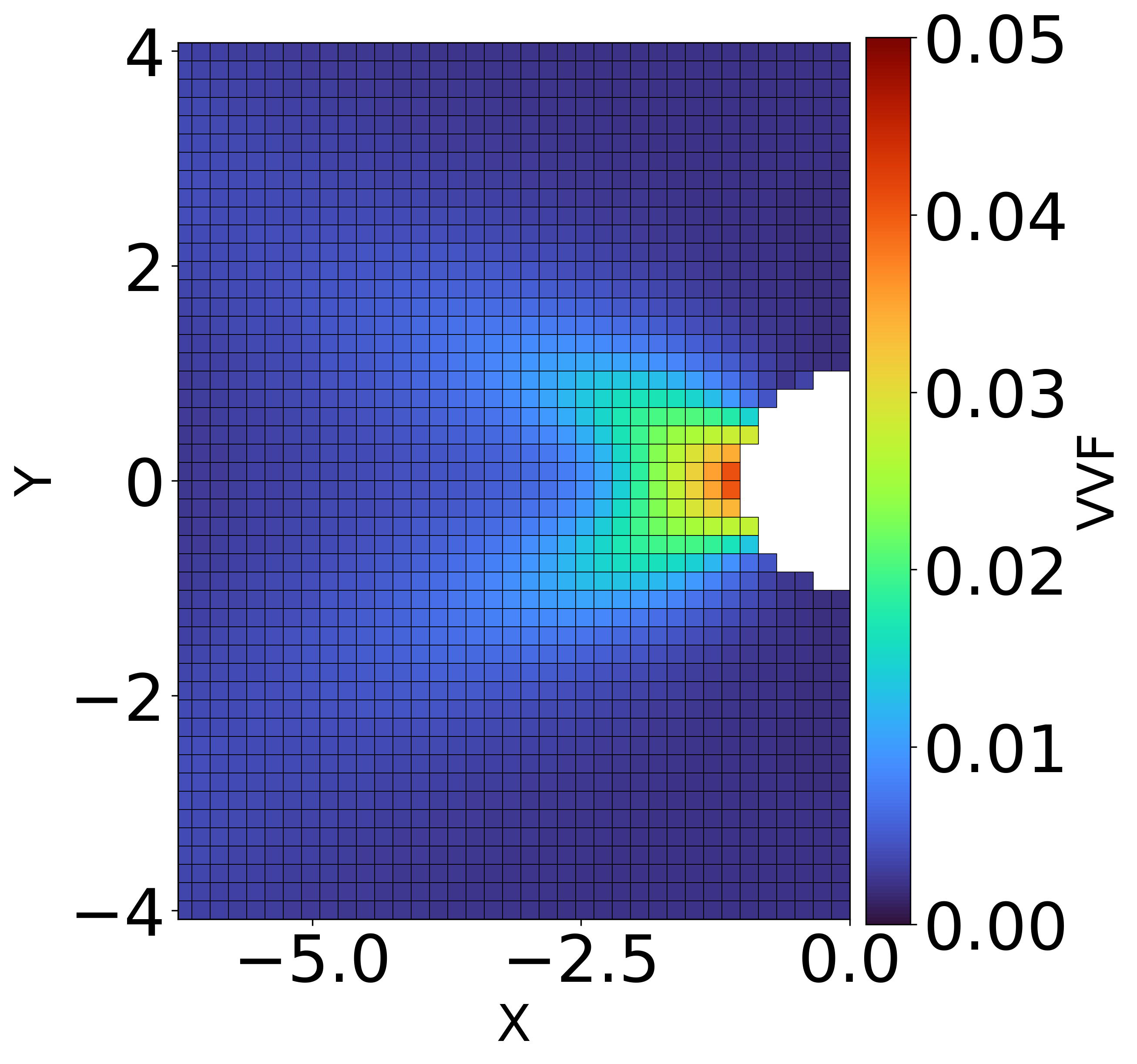}
    \caption{Recovered fields at $F/F_{\max}=0.98$ using the final MAP parameters (F–D$\rightarrow$DIC): (left) axial stress $S_{22}$ [MPa] and (right) void volume fraction (VVF). These fields satisfy both the equilibrium conditions and the estimated GTN material model in a numerical sense.}
    \label{fig:stress_recovery}
\end{figure}

A well-known limitation of DIC is that it provides displacements and strains, but not stresses or damage variables. Prior stress-recovery approaches typically reconstruct stresses from sparse measurements under simplified constitutive assumptions (e.g., linear elasticity, small strain) and rely on regularization to stabilize the inverse problem \citep{Abdollahzadeh2020,Kefal2017}. In contrast, the estimated GTN framework here fuses global F–D and local DIC to identify a nonlinear damage model for the material constitutive law, and can subsequently recover any of the fields computed by the FE model that satisfy the stress equilibrium conditions numerically. We believe that the strategy described above opens a completely new avenue for estimating stress (and other related) fields from the DIC measurements. We briefly demonstrate this capability in Figure~\ref{fig:stress_recovery} by presenting the axial stress field ($S_{22}$) and the void volume fraction (VVF) fields corresponding to the DIC measurement used in the present study (see left column of Figure 12).

This work provides clear evidence of order sensitivity in obtaining posteriors from multi-modal experimental data, as evidenced in the fact that the F–D$\rightarrow$DIC sequence yielded tighter and more stable posteriors than the DIC$\rightarrow$F–D. In a fully specified joint model with exact inference, results should be expected to be order-invariant; however, constructing a single joint likelihood for F–D and full-field DIC is practically infeasible. Therefore, one has to pay particular attention to the sequence in which the posteriors are updated for multi-modal datasets. In our implementation, the observed order sensitivity can be attributed to three likely factors. (1) Finite-sample T-MCMC: the first update gives an approximate posterior from a finite sample set, and reweighting it with the second likelihood can magnify small errors. (2) Same-test dependence: F-D and DIC come from one specimen and one test, so they share test-specific effects (fixture compliance, local geometry/surface, imaging) that our separate noise models may not capture. (3) PCA truncation: to reduce dimensionality, only a few principal components are retained for each dataset, so some information is inevitably lost. The second update operates only on the retained compressed summary; discarded features cannot be recovered. Since the F–D and DIC summaries are learned separately, they emphasize different aspects of the data, so the two different sequences may sharpen different regions of the parameter space. 
The working hypothesis is that the first factor, among the three described, is likely the main contributor. With this in mind, the results are interpreted as follows. Starting with F-D provides a broad, stable picture; sharpening is then achieved by adding DIC, so the F-D$\rightarrow$DIC sequence yields tighter and more stable posteriors. Starting with DIC leaves wider bands that are not substantially reduced by the F–D step. These outcomes reflect practical limitations of the present setup, not a claim that order should matter in principle. 

An important outcome from this study is that we can outline a practically useful protocol for calibration of material models with multi-modal datasets. The recommendation is to first perform single-dataset updates from a uniform prior, quantify each modality’s information content by uncertainty contraction (e.g., HPD-width), and then assimilate in decreasing order of informativeness; the present framework enables this assessment and subsequent sequencing. A more comprehensive study (e.g., a single joint fit, a simple dependence term between F-D and DIC, or short multi-snapshot DIC summaries) might provide further insights. More generally, the single-dataset updates guide the sequence choice: ranking modalities by observed contraction selects F–D before DIC here and, once estimated, also enables recovery of full stress fields consistent with both modalities—an added benefit not available from DIC alone.

\section{Conclusions}
\label{sec:conclusions}

A sequential Bayesian inference (BI) framework was presented for estimating Gurson–Tvergaard–Needleman (GTN) damage parameters from multimodal experiments, integrating specimen-level F–D measurements with spatially resolved DIC strain fields. PCA-based low-dimensional representations for both data streams enabled accurate, efficient Gaussian-process (GP) emulators, making Transitional MCMC (T–MCMC) feasible over a challenging multi-parameter landscape. A voxelized FE mesh aligned simulation outputs with DIC measurements, enabling direct field-level comparisons. In application, multimodal assimilation reconciled global and local observables and materially improved identifiability of damage-evolution parameters. Notably, update order proved consequential: assimilating F–D before DIC concentrated posteriors and yielded parameter sets that simultaneously matched the macroscopic response and post-peak localization patterns.

A key empirical finding is the order sensitivity of the posterior update for multi-modal datasets. Initializing from the stabilizing global response (F–D) and then conditioning on the field data (DIC) produced concentrated and physically consistent posteriors, whereas the reverse ordering yielded broader credible sets. This pattern is plausibly influenced by unequal information content across modalities and same-test dependence, with additional contributions from simplifying approximations (finite-sample T–MCMC and PCA truncation). Practically, the single-dataset updates themselves provide a sequencing rule: from a shared prior, first compute F–D-only and DIC-only posteriors, quantify each modality’s informativeness via uncertainty contraction (e.g., product of univariate posterior 95\% HPD widths or a covariance-determinant proxy), and then assimilate modalities in decreasing order of informativeness. Applied here, this ranking selects F–D before DIC and yields the tightest, most stable GTN posteriors.

Beyond improved identifiability, the estimated GTN parameters enable recovery of stress fields (via the FE model) that are consistent with equilibrium and the constitutive law at the DIC snapshot(s), providing quantities not directly measurable by DIC. These fields support downstream use—e.g., evaluation of triaxiality/Lode indicators and stress-based fracture or fatigue assessments—without additional instrumentation. Compared with elastic inverse reconstructions from sparse measurements, the calibrated GTN framework provides stress fields that are more faithful to the underlying physics, as they are numerically satisfy both equilibrium conditions and the identified inelastic damage model. The framework remains readily extensible: future efforts will incorporate explicit cross-modality dependence in the likelihood, multiple temporal snapshots of the strain fields, and active experiment–simulation design to target the most informative regimes. More broadly, the proposed protocol generalizes to damage models and other multimodal inverse problems in which localized fields complement bulk measurements, offering a practical route to identifiable and uncertainty-aware estimation of complex constitutive models.

\section*{CRediT authorship contribution statement}
\textbf{Mohammad Ali Seyed Mahmoud:} Writing – original draft, Validation, Visualization, Methodology, Experimentation, Simulation, Investigation, Formal analysis, Data curation, Conceptualization. \textbf{Dominic Renner:} Writing – review \& editing, Experimentation, Simulation, Investigation, Formal analysis. \textbf{Ali Khosravani:} Writing – review \& editing, Investigation. \textbf{Surya R. Kalidindi:} Writing – review \& editing, Supervision, Visualization, Funding acquisition, Conceptualization.

\section*{Acknowledgements}
This work was supported by the National Science Foundation (NSF) under grant number 2221104 and by the Army Research Laboratory under Cooperative Agreement Number W911NF-22-2-0106. The views and conclusions contained in this document are those of the authors and should not be interpreted as representing the official policies, either expressed or implied, of the Army Research Laboratory or the U.S. Government. The U.S. Government is authorized to reproduce and distribute reprints for Government purposes notwithstanding any copyright notation herein. The authors gratefully acknowledge the support of the George W. Woodruff School of Mechanical Engineering for providing the necessary facilities and resources.

\section*{Declaration of competing interest}
The authors declare that they have no known competing financial interests or personal relationships that could have appeared to influence the work reported in this paper.

\section*{Data availability}
Data will be made available on request.

\appendix
\section{Score-space likelihood and noise propagation}
\label{app:likelihood}

\subsection{Score-wise Gaussian model}
For each PC score $k$, we model
\begin{equation}
y_k \mid \boldsymbol{\theta} \sim \mathcal{N}\big(\mu_k(\boldsymbol{\theta}),\, v_k(\boldsymbol{\theta})\big),
\qquad
v_k(\boldsymbol{\theta}) \equiv \sigma_k^2 + s_k^2(\boldsymbol{\theta}),
\label{eq:score_model}
\end{equation}
where $\mu_k(\boldsymbol{\theta})$ and $s_k^2(\boldsymbol{\theta})$ are the GP predictive mean and variance for score $k$, and $\sigma_k^2$ is the measurement-noise variance expressed in PC-score space (Section~\ref{app:noise_propagation}). We assume independence across scores.

\subsection{Log-likelihood used in inference}
Under \eqref{eq:score_model}, the (independent) Gaussian log-likelihood is
\begin{equation}
\label{eq:score_loglike}
\mathcal{L}(\boldsymbol{\theta})
= \sum_{k}
\left[
-\tfrac{1}{2}\,\frac{\big(y_k-\mu_k(\boldsymbol{\theta})\big)^2}{v_k(\boldsymbol{\theta})}
-\tfrac{1}{2}\,\log\!\big(2\pi\,v_k(\boldsymbol{\theta})\big)
]
\right.
\end{equation}
and note that $v_k(\boldsymbol{\theta})$ depends on $\boldsymbol{\theta}$ via $s_k^2(\boldsymbol{\theta})$, so the $\sum_k \log v_k(\boldsymbol{\theta})$ term is not a constant. When a constant shift is acceptable, we omit the additive term $-\tfrac{1}{2}\sum_k \log(2\pi)$.

\subsection{Measurement-noise propagation to PC-score space}
\label{app:noise_propagation}
Let the measured vector in the original observation space be
\begin{equation}
\mathbf{z}=\mathbf{y}_{\text{true}}+\boldsymbol{\varepsilon},\qquad
\boldsymbol{\varepsilon}\sim\mathcal{N}(\mathbf{0},\,\Sigma_{\text{meas}}).
\end{equation}
Let $A$ denote the linear preprocessing used prior to PCA (mean-centering and any standardization), and let $\Phi$ be the PC basis matrix (columns are orthonormal PCs). The score vector is
\begin{equation}
\mathbf{s}=\Phi^\top A\,\mathbf{z}.
\end{equation}
By linear propagation of covariance, the noise covariance in score space is
\begin{equation}
\Sigma_s=\Phi^\top A\,\Sigma_{\text{meas}}\,A^\top \Phi,
\end{equation}
and the per-score variances used in \eqref{eq:score_model} are taken from its diagonal:
\begin{equation}
\sigma_k^2=\big[\Sigma_s\big]_{kk}.
\end{equation}

\section{Corner plots for posterior distributions}
\label{app:corner_plots}

This appendix compiles the corner plots for the four runs discussed in Section~\ref{subsubsec:update1}: F-D only, DIC only, F-D$\rightarrow$DIC, and DIC$\rightarrow$F-D. Diagonal panels show 1D marginals; off-diagonal panels show pairwise posteriors.

\paragraph{Key observation}
Across all runs, the clouds are round and axis-aligned, indicating \emph{no visible cross-parameter correlation}. Practical consequence: each parameter is identifiable on its own—tightening one does not require changes in the others—so sequence effects manifest primarily as narrower credible intervals rather than tilted trade-offs.

\begin{figure}[htbp]
  \centering
  \includegraphics[width=0.5\textwidth]{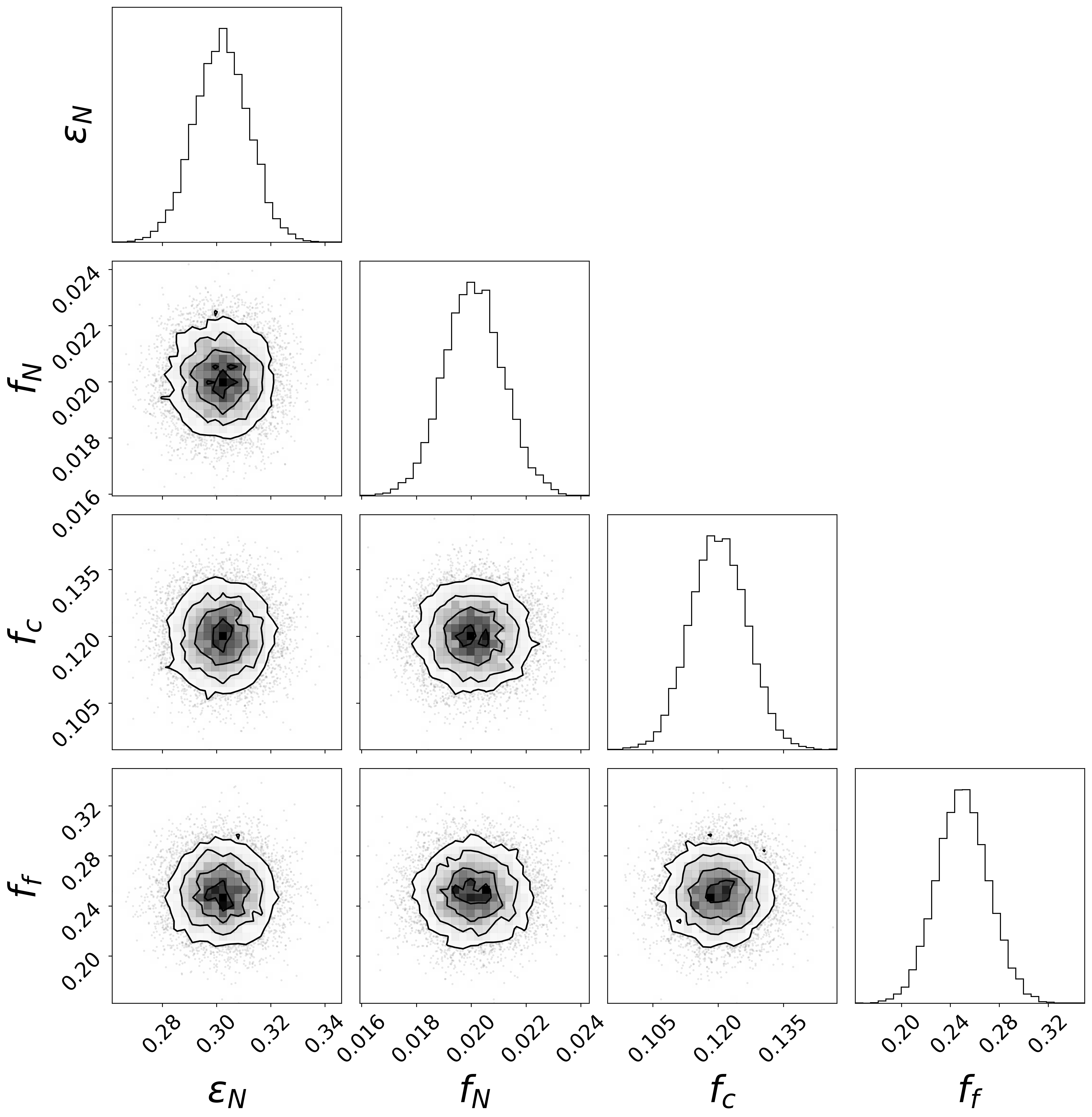}
  \caption{F--D only: corner plot of the posterior over $\boldsymbol{\theta}=\{\varepsilon_N,f_N,f_c,f_f\}$. Unimodal and comparatively concentrated marginals; off-diagonal panels show no apparent parameter correlation.}
  \label{fig:corner_fd}
\end{figure}

\begin{figure}[htbp]
  \centering
  \includegraphics[width=0.5\textwidth]{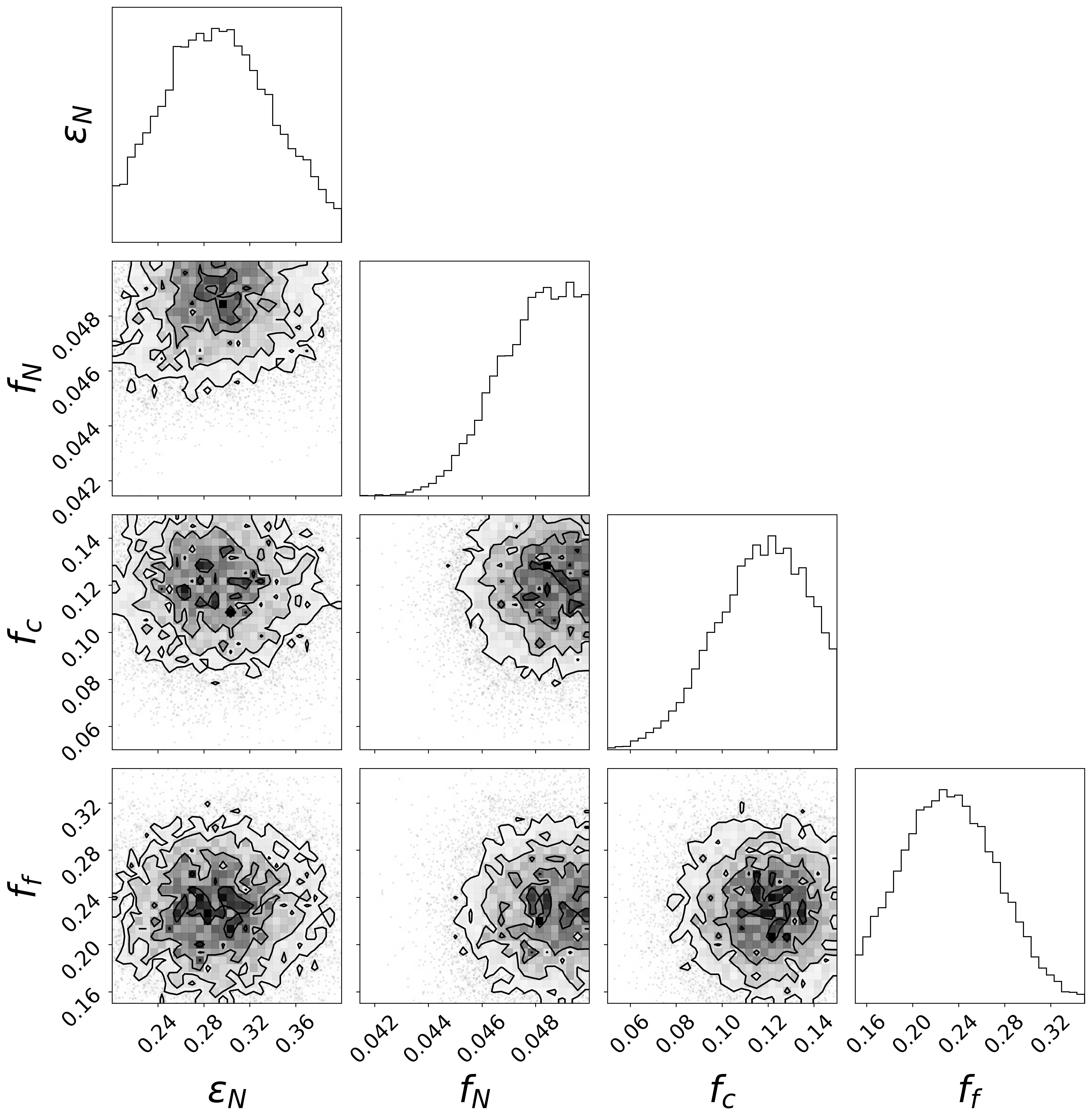}
  \caption{DIC only: broader credible regions with some marginals near prior bounds, indicating limited identifiability under DIC-only data; off-diagonal panels remain approximately axis-aligned.}
  \label{fig:corner_dic}
\end{figure}

\begin{figure}[htbp]
  \centering
  \includegraphics[width=0.5\textwidth]{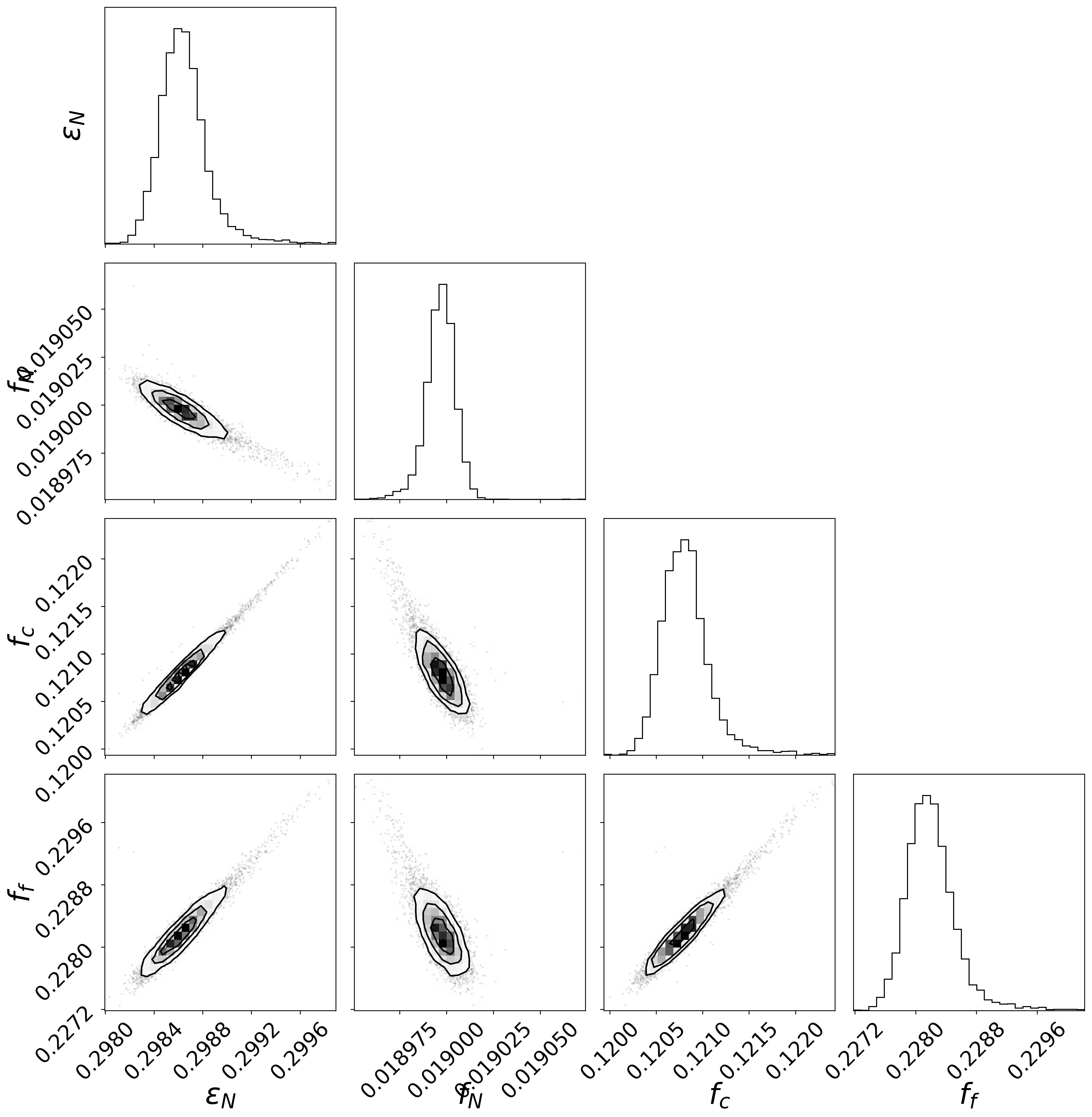}
  \caption{Sequential F--D$\rightarrow$DIC: marked contraction relative to F--D only, especially for $f_N$; pairwise panels remain round/axis-aligned, consistent with negligible cross-parameter correlation.}
  \label{fig:corner_fddic}
\end{figure}

\begin{figure}[htbp]
  \centering
  \includegraphics[width=0.5\textwidth]{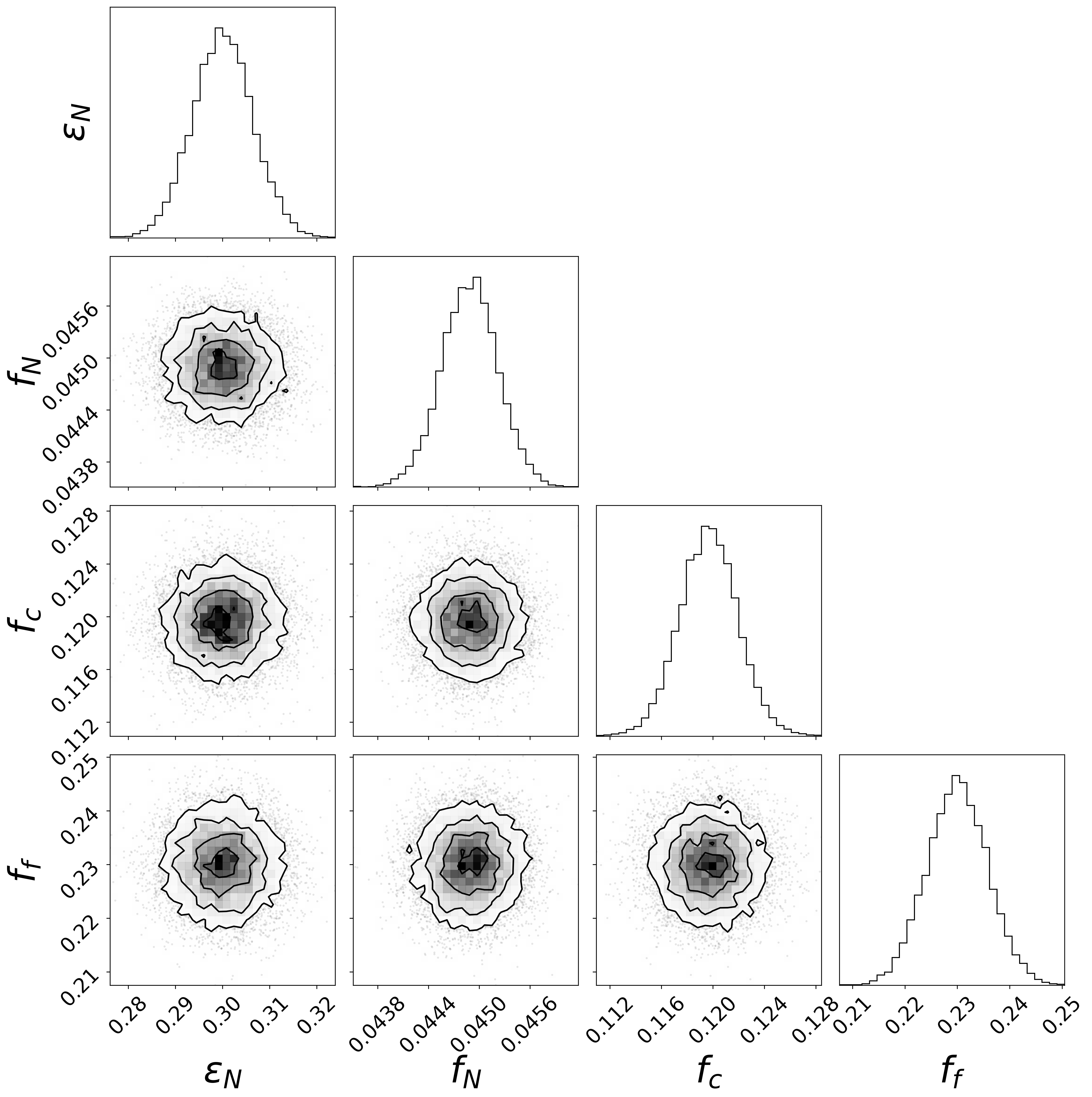}
  \caption{Sequential DIC$\rightarrow$F--D: contraction relative to DIC only, with differences from the F--D$\rightarrow$DIC sequence highlighting order sensitivity; pairwise structure remains nearly uncorrelated.}
  \label{fig:corner_dicfd}
\end{figure}

\end{document}